# Nanocrystals Heterostructures based on Halide Perovskites and Metal Sulfides

Nikolaos Livakas[1,2], Juliette Zito[1], Yurii P. Ivanov[3], Clara Otero Martínez[4], Giorgio Divitini[3*], Ivan Infante[5,6*], Liberato Manna[1*]

[1] Nanochemistry, Istituto Italiano di Tecnologia, Via Morego 30, Genova, Italy

[2] Dipartimento di Chimica e Chimica Industriale, Università di Genova, 16146 Genova, Italy

[3] Electron Spectroscopy and Nanoscopy, Istituto Italiano di Tecnologia, Via Morego 30, Genova, Italy

[4] CINBIO, Department of Physical Chemistry, Materials Chemistry and Physics Group, Universidade de Vigo, Campus Universitario As Lagoas-Marcosende, 36310 Vigo, Spain

[5] BCMaterials, Basque Center for Materials, Applications, and Nanostructures, UPV/EHU Science Park, Leioa 48940, Spain

[6] Ikerbasque Basque Foundation for Science Bilbao 48009, Spain

**ABSTRACT:** We report the synthesis of nanocrystal heterostructures composed of $CsPbCl_3$ and $PbS$ domains sharing an epitaxial interface. We were able to promote the growth of a $PbS$ domain (in competition with the more commonly observed $Pb_4S_3Cl_2$ one) on top of the $CsPbCl_3$ domain by employing $Mn^{2+}$ ions, the latter acting most likely as scavengers of $Cl^-$ ions. Complete suppression of the $Pb_4S_3Cl_2$ domain growth was then achieved by additionally selecting an appropriate sulfur source (bis(trimethylsilyl)sulfide, which also acted as scavenger of $Cl^-$ ions), and reaction temperature. In the heterostructures, emission from the perovskite domain was quenched, while emission from the $PbS$ domain was observed, pointing to a type-I band alignment, as confirmed by calculations. These heterostructures in turn could be exploited to prepare second-generation heterostructures through selective ion exchange on the individual domains (halide ion exchange on $CsPbCl_3$, cation exchange on $PbS$). We demonstrate the cases of $Cl^- \rightarrow Br^-$ and $Pb^{2+} \rightarrow Cu^+$ exchanges, which deliver $CsPbBr_3/PbS$ and $CsPbCl_3/Cu_{2-x}S$ epitaxial heterostructures, respectively.

## INTRODUCTION

Colloidal semiconductor nanocrystals (NCs) of all-inorganic lead halide perovskites (LHP) ($CsPbX_3$, $X = Cl^-$, $Br^-$, $I^-$) have attracted a broad research interest owing to their notable optoelectronic properties, such as high absorption coefficient and photoluminescence quantum yield (PLQY), narrow emission bandwidth, and defect tolerance.[1-5] More recently, research has also expanded in the direction of colloidal NC heterostructures, in which one domain is a halide perovskite.[6-23] Traditionally, heterostructures with heterodimer or core-shell geometries have been proven beneficial in terms of stability,[24] bandgap tunability,[25, 26] and the emergence of new properties,[27] leading to their extensive investigation in light-emitting devices,[28, 29] catalysis,[28] and biomedicine.[30] For instance, in a core/shell geometry with a type-I band alignment, the shell generally improves the stability and PLQY of the light emitted from the core,[31] whereas in a heterodimer geometry in which the two domains have a type-II band alignment, opposite charge carriers can be separated across the heterojunction and can then be individually exploited for catalytic applications.[24, 32]

To date, various materials have been coupled to LHPs to form heterostructures or core/shell structures. The most studied cases are metal oxides ($SiO_2$, $TiO_2$),[33-36] metals (Au, Pt, Bi),[17-20] metal halides ($Cs_4PbBr_6$, $CsBr$),[11, 15, 16] chalcogenides (CdS, PbS),[10, 11, 13, 14, 37-39] dichalcogenides ($MoSe_2$),[40] and chalcohalides ($Pb_4S_3Br_2$).[9, 12, 23] Only in a subset of these works was the interface between the LHP and the second material studied in detail and found to be epitaxial. A case study of an epitaxial interface was reported by our group, involving $CsPbBr_3$ and the chalcohalide $Pb_4S_3Br_2$.[23] In it, the emission from the LHP domain was quenched by the presence of the chalcohalide domain. In a later work, those heterostructures were found to promote photocatalytic $CO_2$ reduction.[12] Research in this direction was extended by us to the $CsPbCl_3/Pb_4S_3Cl_2$ case.[9] Another recent example of heteroepitaxy is the one between $CsPbBr_3$ and CdS, a type-II heterojunction providing an improved sensitivity in photodetection.[11] An example of a type-I alignment, and the focus of this work, is found in $CsPbBr_3$-PbS. Although in previous works there have been attempts to grow this type of heterostructure and study its

properties, no clear evidence was presented of an epitaxial connection between the two domains.[13, 39]

We report here a synthesis route to CsPbCl$_3$/PbS epitaxial NC heterostructures. Our approach is similar to the one previously developed by us to grow CsPbCl$_3$/Pb$_4$S$_3$Cl$_2$ heterostructures:[9] in that case, preformed CsPbCl$_3$ clusters (Scheme 1, "synthesis of CsPbCl$_3$ clusters") were injected in a reaction environment that promotes their evolution to larger CsPbCl$_3$ NCs, followed by the heterogenous nucleation of a Pb$_4$S$_3$Cl$_2$ domain on them (Scheme 1, "Synthesis of Heterostructures", Case I). This latter domain grows using a fraction of Pb$^{2+}$ and Cl$^-$ ions derived from the decomposition of the initial CsPbCl$_3$ clusters, plus a sulfur source. In the present work, we realized that to selectively promote the heterogeneous nucleation of PbS over that of Pb$_4$S$_3$Cl$_2$ (thus forming CsPbCl$_3$/PbS heterostructures instead of CsPbCl$_3$/Pb$_4$S$_3$Cl$_2$ ones) we had to reduce the availability of the Cl$^-$ species in solution. This was made possible by the addition to the reaction environment of an exogenous cation, Mn$^{2+}$, which does not participate in the crystallization processes, yet, likely by binding to Cl$^-$ ions in the solution phase, reduces their availability and thus promotes the heterogeneous nucleation of PbS instead of Pb$_4$S$_3$Cl$_2$ (Scheme 1, Cases II-IV).

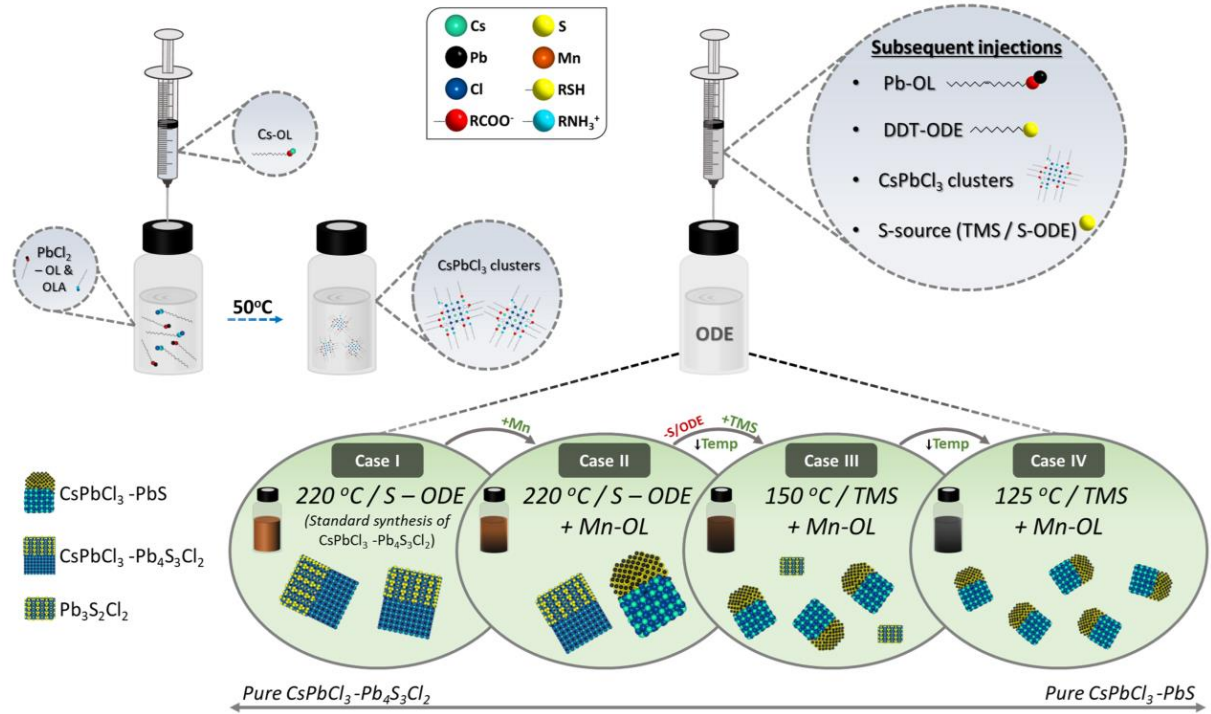

**Scheme 1. Heterostructures synthesis.** Step 1: Synthesis of CsPbCl$_3$ nanoclusters; Step 2: synthesis of heterostructures, by mixing the CsPbCl$_3$ nanoclusters with different reactants, under various reaction conditions. Case I: our previously published protocol, leading to CsPbCl$_3$/Pb$_4$S$_3$Cl$_2$ heterostructures.[9] Case II: including Mn-oleate promotes the growth of CsPbCl$_3$/PbS in addition to CsPbCl$_3$/Pb$_4$S$_3$Cl$_2$. Case III: replacing S-ODE with a more reactive sulfur source (TMS) and working at a lower temperature (150 °C) leads to CsPbCl$_3$/PbS as the only heterostructures, albeit smaller than in case II, and with isolated chalcohalides as a byproduct. Case IV: CsPbCl$_3$/PbS heterostructures are obtained as the only product at even lower temperatures (125 °C), with an overall size smaller than in case III.

We found that there is a strong competition between the chalcogenide (PbS) and chalcohalide (Pb$_4$S$_3$Cl$_2$) phases to heterogeneously nucleate on the CsPbCl$_3$ domain, to the point that there is a narrow window of experimental parameters in which PbS prevails. The growth of CsPbCl$_3$/PbS heterostructures carrying CsPbCl$_3$ and PbS domains that are large enough (15 nm or bigger) for an accurate characterization of the epitaxial interface by transmission electron microscopy (TEM) comes at the cost of a conspicuous co-presence of CsPbCl$_3$/Pb$_4$S$_3$Cl$_2$ heterostructures (Scheme 1, Case II). Instead, purer, although smaller-sized CsPbCl$_3$/PbS heterostructures were grown by using a more reactive sulfur precursor (bis(trimethylsilyl)sulfide), which could also scavenge Cl$^-$ ions through its reactive trimethylsilyl groups, and working at comparatively lower temperatures (125°C-150°C instead of 220°C, Scheme 1, Cases III-IV). Scanning transmission electron microscopy (STEM) analysis indicated a well-defined epitaxial interface between the perovskite and lead sulfide domains, due to the continuity of the Pb sublattice across the interface between the two domains: they share a common Pb layer at the junction, which coordinates with Cl$^-$ ions on one side and S$^{2-}$ ions on the other side. Density functional theory (DFT) calculations suggested a quasi-type-I band alignment at the heterojunction, with band-edge electrons fully localized in the PbS domain and holes that are instead localized in a region at the interface of the two domains, spanning both of them. These calculations are in accordance with the optical emission spectra of the heterostructures exhibiting PL in the near-infrared (NIR) region (originating from PbS), but weak PL in the blue region (originating from $CsPbCl_3$), indicative of predominant carrier transfer from the $CsPbCl_3$ to the PbS domain upon photoexcitation.

Our samples could then be used as precursors for the synthesis of second-generation heterostructures by exploiting the selectivity of the two domains to different ion exchange reactions. For example, by exchanging $Pb^{2+}$ ions with $Cu^+$ ions we could transform the PbS domain to $Cu_{2-x}S$, delivering $CsPbCl_3$-$Cu_{2-x}S$ heterostructures. Alternatively, by exchanging $Cl^-$ ions with $Br^-$ ions we could transform the $CsPbCl_3$ domain to $CsPbBr_3$, delivering $CsPbCl_3$-$PbS$ heterostructures. In both cases, the epitaxial connection between the ion-exchanged and the unexchanged domain was maintained. Our approach demonstrates a flexible strategy to synthesize NC heterostructures in which one of the two domains is a halide perovskite and the other is a metal chalcogenide.

## RESULTS AND DISCUSSION

The synthesis of $CsPbCl_3$/PbS colloidal heterostructure NCs was carried out via the cluster-based approach (Scheme 1). This is a two-step method that has been established as an effective route for the synthesis of several perovskite-based heterostructures.[9, 10, 23] The first step in this method includes the synthesis of $CsPbCl_3$ nanoclusters at a relatively low temperature (50 °C) over a rather long reaction time (30 min), made possible by a high concentration of surfactants, namely oleic acid and oleylamine (see details in the experimental section).[41] These clusters are then isolated, purified, and resuspended in octadecene (ODE), to be used as precursors in the second step. This latter step is a modification of our previously developed synthesis of $CsPbCl_3$/$Pb_4S_3Cl_2$ heterostructures and consists of sequential injections, into degassed ODE at high temperature (220 °C), of Pb-oleate, dodecanethiol, preformed $CsPbCl_3$ nanoclusters, and elemental sulfur (dissolved in ODE, henceforth referred to as S-ODE).[23]

In this work, the only additional ingredient is represented by $Mn^{2+}$ ions, introduced as Mn-oleate. Initially, the feed ratio $X_{Mn}$ of Mn (defined as $X_{Mn}$ = [Mn]/([Mn]+[Pb])*100) was 33%. This synthesis led to a mixture of the usual $CsPbCl_3$/$Pb_4S_3Cl_2$ heterostructures along with larger, mushroom-like heterostructures (Figure 1a), which were identified as $CsPbCl_3$/PbS based on high-resolution scanning TEM energy dispersive X-Ray Spectroscopy (HR STEM-EDX, Figure 1d). The coexistence of these two populations of heterostructures was further confirmed through X-ray Diffraction (XRD, Figure 1f,g). The presence of $Mn^{2+}$ was instrumental in the competitive formation of $CsPbCl_3$/PbS, but only trace amounts of Mn were actually found in the NCs, and mainly in the PbS domain (see STEM-EDX elemental mapping in Figure S1 and the corresponding analysis in Table S1). We therefore concluded that $Mn^{2+}$ ions influence the kinetics of heterogeneous nucleation of PbS and $Pb_4S_3Cl_2$, most likely by complexing a fraction of the available $Cl^-$ ions in solution. This is supported by the stronger bond dissociation energy of Mn-Cl (361 kJ/mol) compared to Pb-Cl (301 kJ/mol),[42] and our DFT calculations of the binding affinity of $Cl^-$ ions to $Pb^{2+}$ and $Mn^{2+}$ ions (-384.76 kcal/mol for $PbCl_2$ and -459.81 kcal/mol for $MnCl_2$, see experimental section). Also, Pb binds stronger to S (346 kJ/mol) than to Cl (301 kJ/mol), plus the formation of PbS over any Mn-based compounds might favored by the lower lattice mismatch of PbS with $CsPbCl_3$.[43] We also tested $Na^+$, in the form of metal oleate (see Figures S2, S3), since the Na-Cl bond also has strong dissociation energy (410 kJ/mol).[42] In this case, we also observed the formation of $CsPbCl_3$/PbS heterostructures, similar to the $Mn^{2+}$ case, however, accompanied by isolated NCs of various compositions, including large $CsPbCl_3$ NCs and amorphous aggregates containing mainly Na and Cl (based on EDS elemental mapping, Figure S3). Hence, also $Na^+$ ions are scavenging $Cl^-$ ions, but they do not seem to work as effectively as $Mn^{2+}$ ions in terms of purity of the final product. We also need to stress that other parameters might play important roles in these reactions, such as the stabilities of the various metal-ligand complexes and the lattice energies of the various phases that are formed. All these roles are hard to disentangle.

In this sample grown using S-ODE, a 33% Mn feed ratio, and a 220° reaction temperature, the $CsPbCl_3$/$Pb_4S_3Cl_2$ heterostructures were the predominant product, as evidenced by XRD, optical spectra, and TEM analysis (Figure 1e,f,g, and S4a). The modification of this synthesis scheme, for example by increasing the $Mn^{2+}$ feed ratio ($X_{Mn}$ > 33.3%) did not lead to an increase in the population of PbS-based heterostructures (Figure S4) but rather to a mixture of isolated chalcohalide NCs, PbS NCs, and non-uniform heterostructures (both $CsPbCl_3$/$Pb_4S_3Cl_2$ and $CsPbCl_3$/PbS), as indicated by XRD (Figure S4e). Different reaction temperatures (spanning the 135°- 250°C window) had an impact on the size of the heterostructures, but not on their phase purity (Figures S5-S7), and indeed the XRD patterns continued to show the presence of mainly the $Pb_4S_3Cl_2$ phase. Also, the sample showed no IR emission from PbS, most likely because the PbS domains were quite large. This unfortunately betrayed our expectations of being able to promote the almost exclusive nucleation of PbS at lower temperatures, based on our considerations on the lower formation energy of lead sulfide over that of lead chalcohalides.[44]

Switching to a more reactive sulfur source than S-ODE, namely bis(trimethylsilyl)sulfide (TMS, see Figures 1b,c and S8-S13), led instead to phase-pure $CsPbCl_3$/PbS heterostructures in the 125 °C-200 °C temperature window. At low temperatures (125 °C), the heterostructures were small in size (~7.2 nm) and were emitting in the NIR range (with a peak at 1250 nm, see Figure 1e and S9), indicating the formation of quantum confined PbS. The phase purity and the dominance of PbS over $Pb_4S_3Cl_2$ were evident from XRD (Figures 1f,g, and S9b), high-angle annular dark field (HAADF) HR STEM, and STEM-EDX (Figure S10). Bigger $CsPbCl_3$/PbS heterostructures could be synthesized at higher reaction temperatures (Figures S8-S9a). However, at reaction temperatures exceeding 140 °C, small and isolated $Pb_3S_2Cl_2$ NCs were also formed. These could easily be removed by size-selective precipitation (Figure S11), thus recovering pure $CsPbCl_3$/PbS heterostructures (Figure 1b).

At higher temperatures (>160 °C), the heterostructures were more heterogeneous in size/shape (Figure S8), and their PbS-CsPbCl$_3$ purity (after size-selective precipitation) remained until 180 °C, as confirmed by XRD (at 200 °C the reflections of the Pb$_4$S$_3$Cl$_2$ phase started to appear).

Reaction time was also an important parameter in controlling the growth of the heterostructures. Figures S12 and S13 demonstrate the heterostructure's size tunability for two different time intervals (5 and 20 minutes) at three reaction temperatures (135 °C, 140 °C, and 150 °C). In all cases, we observed a gradual increase in heterostructure size. The aforementioned effective strategy (150 °C/TMS) to monodispersed CsPbCl$_3$/PbS heterostructures may also stem from the ability of the trimethylsilyl group to act as an efficient scavenger for Cl$^-$ ions. This should occur through a dehalosilylation reaction that would produce a stable and volatile trimethylsilylchloride.[45] We verified that TMS alone, in the absence of Mn$^{2+}$, can also lead to CsPbCl$_3$/PbS heterostructures, albeit contaminated with other products (Figure S14). However, the synergistic use of an exogenous cation (Mn$^{2+}$) and TMS has proven to be the most effective approach to prepare pure PbS-based heterostructures. It is worth of note that the emission from the perovskite domain was severely quenched when it was bound to a PbS domain (Figure S15). This could be attributed to the band alignment at the heterojunction, as will be discussed later.

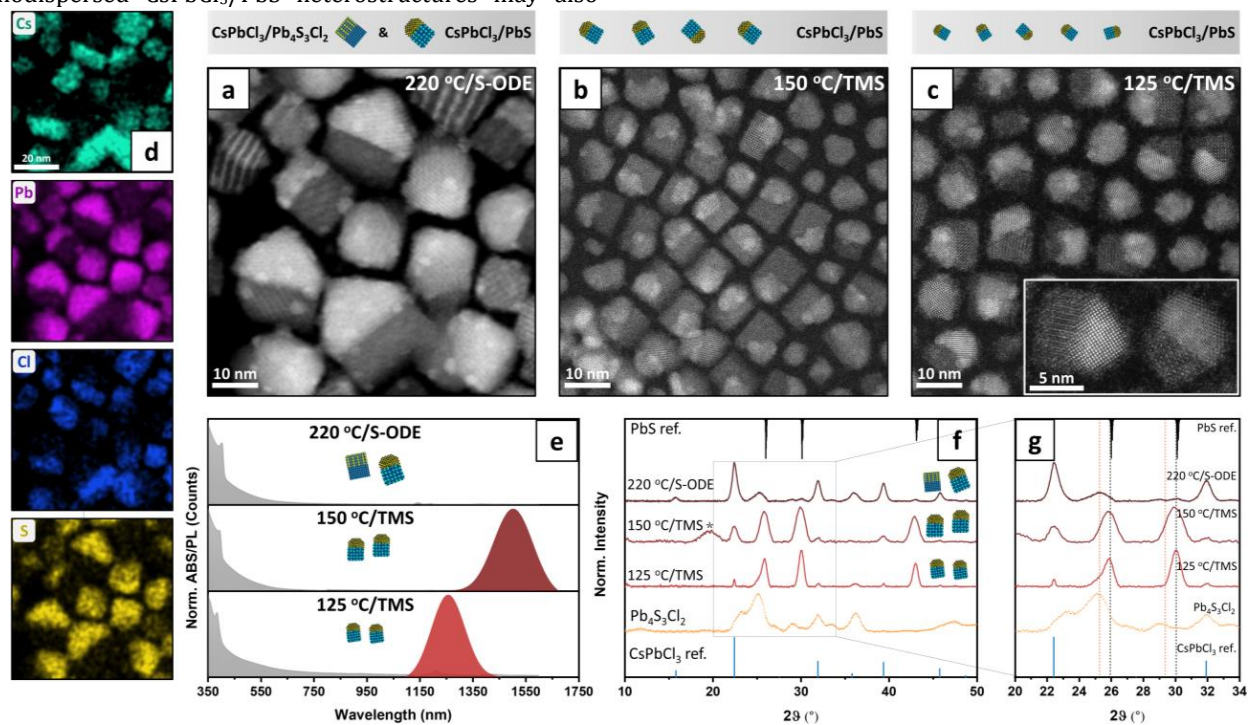

**Figure 1. Analysis of heterostructures from cases II, III, and IV of Scheme 1**. (a-c) STEM-HAADF images of each sample. (d) STEM-EDX elemental maps of the same area in (a) demonstrate the coexistence of two types of heterostructure in this sample. (e) Optical absorption (grey) and PL (colored) spectra of the three samples (the 220 °C/S-ODE" does not have any NIR emission). f) XRD patterns of the three samples and Pb$_4$S$_3$Cl$_2$, highlighting the CsPbCl$_3$/PbS phase purity for the "125 °C/TMS" and "150 °C/TMS" samples, and the coexistence of the two types of heterostructures in "220 °C/S-ODE". The peak at 20° (marked with *), is attributed to fumed silica used to prepare the XRD samples. (g) Magnified view of the same XRD patterns in the 20°-34° region, highlighting the presence of pure PbS phase in the "125 °C/TMS", and "150 °C/TMS" samples, with distinctive diffraction peaks at 26° and 30° (black dashed lines). The "220 °C/S-ODE" sample presents a dominant chalcohalide-based composition, as indicated by the diffraction peaks at 25.2°, and 29.1° (orange dashed lines).

The epitaxial relationships between the two domains of the CsPbCl$_3$/PbS heterostructure were investigated by HAADF STEM. Figure 2 reports atomic resolution images of single heterostructures where the CsPbCl$_3$ and PbS domains share a sharp interface. Several projected views of the heterostructure are accessible: for example, two different CsPbCl$_3$/PbS NCs are shown here (Figure 2b,c) in mutually orthogonal projections. In one of them (Figure 2b), both domains present low-index zone axes ([-110] for PbS and [00-1] for CsPbCl$_3$), and the atomic columns are visible, enabling atomic resolution imaging of the interface. In the other projection (Figure 2c), only the perovskite domain is on a low-index zone axis [0-10], and the atomic columns in PbS domain are not clearly resolved, as it is viewed from a high-index direction. However, by tilting the sample by 7° around an axis laying in the horizontal direction, the [00-2] zone axis could be accessed (Figure 2d), allowing us to identify the periodicity and rock-salt cubic crystal structure of the PbS domain. The mutual orthogonality of the projections displayed in Figure 2b and 2c was further confirmed by the FFT analysis of the corresponding STEM images (Figures 2e,f, and S16). Using the extracted crystallographic information, we constructed a 3D model of the heterostructure. This model, reported in Figure 2h, closely matches the atomic columns of each heterostructure projection, as represented in the magnified images (Figures

2a, and S16). A closer look at the interface revealed a stair-like repetitive motif between the two domains, as depicted in Figure 2a,i. Such an interface exposes multiple sets of parallel planes at the junction. The interface plane is depicted by a dashed red line in Figure 2a,h,i. It consists of the (-4-42) PbS and (-210) CsPbCl$_3$ sets of parallel planes of the respective domains. The directions of these crystallographic planes can be inferred from the corresponding FFT patterns (Figures 2e,f) illustrating their parallel alignment along the interface. Figures 2i,j are magnified and rotated views of the interface, so that is it either seen in side-view (Figure 2i) or top-view (Figure 2j, that is, along the vector perpendicular to the interfacial plane).

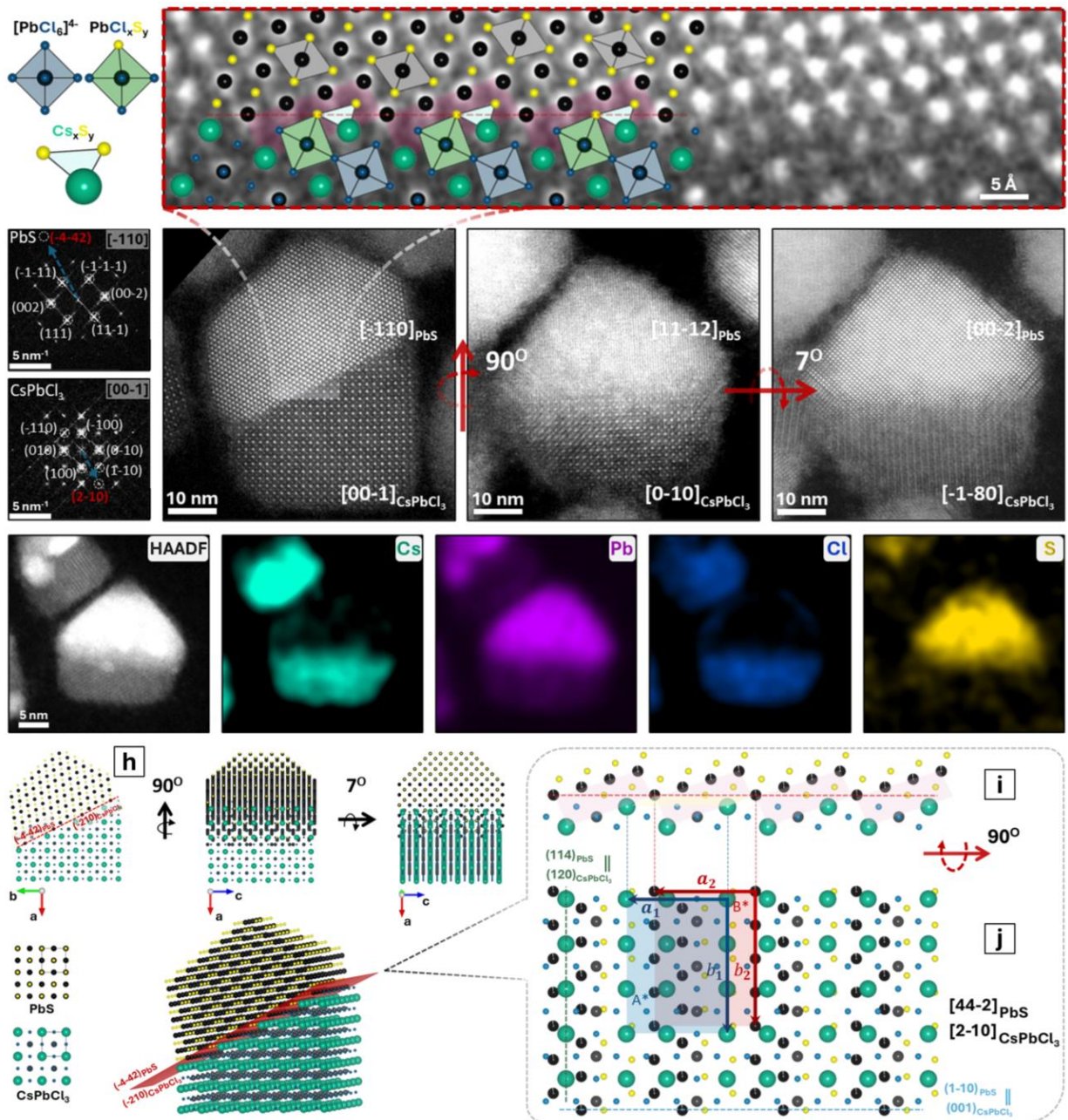

**Figure 2. Structural and compositional analysis of CsPbCl$_3$/PbS epitaxial heterostructures.** (a) 3D reconstructed model superimposed on the magnified view of the stair-like epitaxial interface between CsPbCl$_3$ and PbS domains (the corresponding HAADF STEM image is reported in panel b). The [PbCl$_6$]$^{4-}$ octahedra in the bulk of the perovskite domain are in light blue, while the PbCl$_x$S$_y$ octahedra at the interface are in light green, underlining the continuity of the Pb$^{2+}$ sublattice through coordination with S$^{2-}$ ions. The dashed red line crossing the Pb$^{2+}$ ions at the interface corresponds to the (-4-42) PbS and (-210) CsPbCl$_3$ sets of parallel planes. (b-d) HAADF STEM images of two heterostructures in mutually orthogonal orientations. Panel (d) is the same heterostructure as in (c) after a 7° tilt around an axis lying in the horizontal direction. The orthogonal projection is represented in (b). In this case, the heterostructure aligns both domains with low-index zone axes ([-110] for PbS and [00-1] for CsPbCl$_3$). (e,f) FFT patterns of the corresponding PbS (e) and CsPbCl$_3$ (f) domains of the panel (b). Red dashed circles represent two sets of planes

parallel to the interface. g) HAADF STEM image of a single heterostructure with the corresponding STEM-EDX elemental maps. (h) 3D model of CsPbCl$_3$/PbS heterostructures shown in the same orientations as the heterostructures in panels (b,c,d). (i) cross-sectional view of the interface plane. (j) Top view of the interface plane between the two domains (90º rotation around x axis from the cross-section view presented in (i)). Two distinct surface supercells A* and B* are identified (red and blue rectangles) for each domain, indicating minimal lattice mismatch (<0.4 % for a and b).

We marked two distinct surface supercells A* and B* (red and blue rectangles) and their corresponding lattice vectors $\vec{a_1} = \alpha_1 \vec{g}_{[120]_{CsPbCl3}}$, $\vec{a_2} = \alpha_2 \vec{g}_{[114]_{PbS}}$, and $\vec{b_1} = b_1\vec{g}_{[001]_{CsPbCl3}}$, $\vec{b_2} = b_2\vec{g}_{[1-10]_{PbS}}$, where g is the unit vector of the corresponding crystallographic direction, for each crystal. These are repetitive two-dimensional unit cells of each phase that can reproduce the crystal lattice at the interface by translation. The slight difference in dimensions of the two cells suggests a minimal lattice mismatch (less than 0.4 % for *a* and *b*), which is a crucial factor for isostructural phases to establish a favorable epitaxial relationship. Elemental analysis by STEM-EDX confirmed a homogeneous distribution of the constituent elements within the two distinct domains of a single heterostructure (Figure 2g) as well as its stoichiometry.

Aided by the microscopy analysis discussed above, we performed atomistic calculations at the DFT level of theory to better understand the structural features of the CsPbCl$_3$/PbS heterostructures, particularly focusing on the interface between the two domains. We started by preparing two separate models of CsPbCl$_3$ and PbS NCs in shapes that would match the two separate domains, if they were allowed to grow individually as single NCs, i.e. respectively as a cube for CsPbCl$_3$ and a cuboctahedron for PbS. These are illustrated in Figure 3a. The cubic CsPbCl$_3$ NC model of about 3.4nm on each side was essentially the same as the CsPbBr$_3$ NC models widely reported in the literature, and exhibiting six (001) facets.[46, 47] The structure of this model was relaxed at the DFT level. Concomitantly, a PbS NC model of about 3.0 nm was prepared by carving a bulk rock-salt structure to prepare a NC featuring six stoichiometric (100) facets and eight anion-rich (111) facets. In this case, the PbS structure was relaxed at the DFT level as well. In both models, Cl$^-$ anions were employed as a convenient way to simulate surface ligands while containing the computational effort and preventing complications in the analysis. As illustrated in Figure 3a, to prepare the heterostructure we first oriented the CsPbCl$_3$ model along the [001] axis and cut the (110) planes in a stair-like fashion to reproduce the experimental pattern of Figure 2b-d. We then aligned the PbS NC model along the [110] axis and cut the (111) planes in a stair-like fashion (with a repetition unit of four Pb cations) to match the experimental STEM image of Figure 2b. Finally, we made the two fragments fit together to obtain a CsPbCl$_3$/PbS NC model featuring a charge-balanced Cs$_{126}$Pb$_{337}$S$_{221}$Cl$_{358}$ stoichiometry, as depicted in Figure 3b. Importantly, the stair-like epitaxial alignment enables complete coordination of the Pb$^{2+}$ cations at the interface of the PbS and CsPbCl$_3$ domains, where the anions (either Cl$^-$ or S$^{2-}$) fill the halide vacancies of the perovskite fragment. This compensation results in a heterostructure without overall halide vacancies, leading to a smooth relaxation of the final model at the DFT level.

Based on this model, we computed the electronic structure and performed an inverse participation ratio (IPR) analysis to evidence the presence of localized states in the electronic structure (IPR values deviating significantly from zero), as presented in Figure 3c.[48] The computed heterostructure features a clean band gap, free of mid-gap states: here, the conduction band edge states are mainly delocalized over the PbS domain, whereas the valence-band edge states, also dominated by the contribution of the PbS domain, present intermixed contributions of the two domains, and are more localized at the interface.

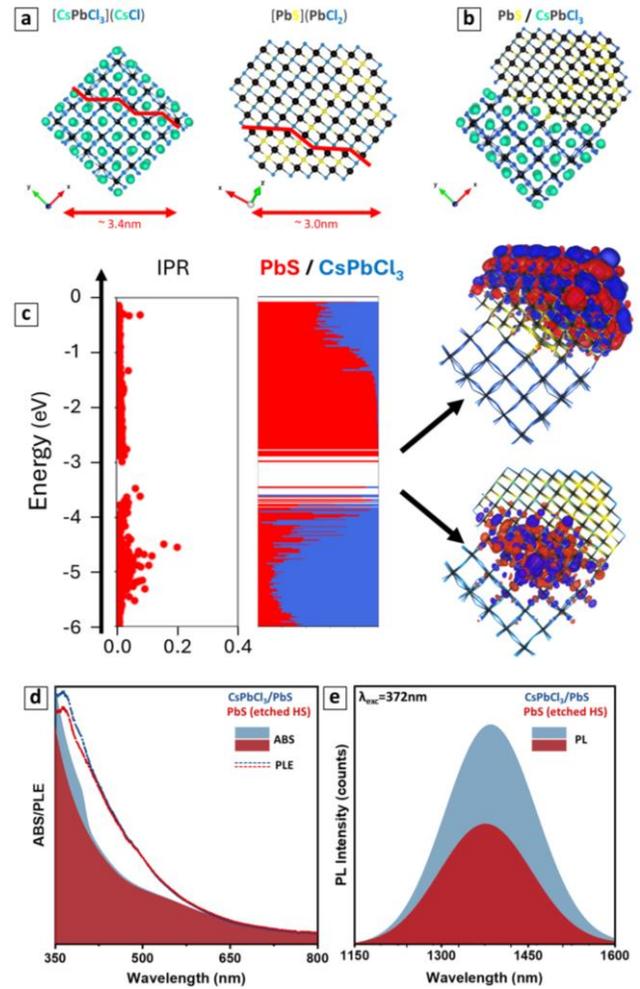

**Figure 3. Computational and optical properties.** (a) Ball and sticks representation of (left) a cubic ~3.4 nm-sided CsPbCl$_3$ NC model and (right) a cuboctahedral ~3.0 nm-sided PbS NC model optimized at the DFT/PBE level of theory. (b) Preparation of the heterostructure NC model by cutting the PbS and CsPbCl$_3$ models along the planes evidenced as red lines in (a) and stacking along the [110] axis to fit the experimental STEM images. (c) IPR plot and electronic structure of the CsPbCl$_3$/PbS NC heterostructure model computed at the DFT/PBE level of theory. The color code

indicates the contribution of each domain to each molecular orbital. On the right are plotted the frontier molecular orbitals. (d) Optical absorption (colored curves) and PL excitation (dashed lines) spectra for both CsPbCl$_3$/PbS heterostructures and PbS NCs recovered after etching away the CsPbCl$_3$ domain. (e) PL spectra of the same samples, using an excitation wavelength at 370 nm.

We also directly compared the optical properties of CsPbCl$_3$/PbS heterostructures with those of the PbS NCs recovered from them after etching away the CsPbCl$_3$ domain (see Experimental Section for details). XRD and TEM analysis of this sample indicated pure PbS (Figures S17 and S18). The absorption spectra of the two samples are almost superimposable, except for the feature in the UV region for the heterostructure sample, ascribed to the perovskite domain (Figure 3d). The PL in the NIR region from the PbS sample was less intense and slightly blue-shifted compared to the heterostructure sample (Figure 3e, both samples had the same optical density at 500 nm). We may attribute the blue shift either to a slightly smaller PbS size resulting from the removal of surface layers during etching (Figure S18) or to a slightly stronger confinement of the carriers in free-standing PbS compared to PbS interfaced with CsPbCl$_3$. Photoluminescence excitation (PLE) spectra for both samples (recorded at 1380 nm, Figure 3d) closely resembled their respective absorption curves. This, for the heterostructure, indicates that the NIR emission may also originate from the carriers photogenerated in the CsPbCl$_3$ domain and transferred from there to the PbS domain.

Post-synthesis ion exchange reactions in colloidal NCs have been established as an effective strategy to access a wide range of nanomaterials, including alloys or heterostructures, which are challenging to prepare with direct synthesis methods.[49-52] In a similar fashion, the CsPbCl$_3$/PbS heterostructures may be used as templates for post-synthesis ion exchange reactions in both domains of the dimer. This would enable selective modification of the composition of each domain. As a demonstration of this versatile process, we exposed the CsPbCl$_3$/PbS heterostructures to either Br$^-$ or Cu$^+$ ions. This allowed us to induce either full anion exchange in the perovskite domain or full cation exchange in the metal sulfide domain, resulting in the formation of two possible daughter heterostructures, CsPbBr$_3$/PbS, and CsPbCl$_3$/Cu$_{2-x}$S. Such exchanges were carried out using two different samples of CsPbCl$_3$/PbS heterostructure, i.e. those previously named "150 °C/TMS" and "220 °C/S-ODE" (see Figure 1). The former sample, as previously discussed, consisted of small, pure CsPbCl$_3$/PbS heterostructure, and allowed an easier investigation of the crystal phase (Figure S19) and optical properties (Figures 5c and S20) after the exchange. The latter sample consisted instead of bigger (albeit less pure) CsPbCl$_3$/PbS heterostructures, to facilitate their microscopy study (Figure 4, S21, and S23). Starting from the "150 °C/TMS" sample, both exchanges were confirmed by modifications in the optical spectra and XRD patterns (Figures S19, 5c, and S20). The Br-exchanged sample presented the characteristic excitonic absorption of the CsPbBr$_3$ perovskite phase while maintaining the IR absorption of the PbS domain (Figure 5c).

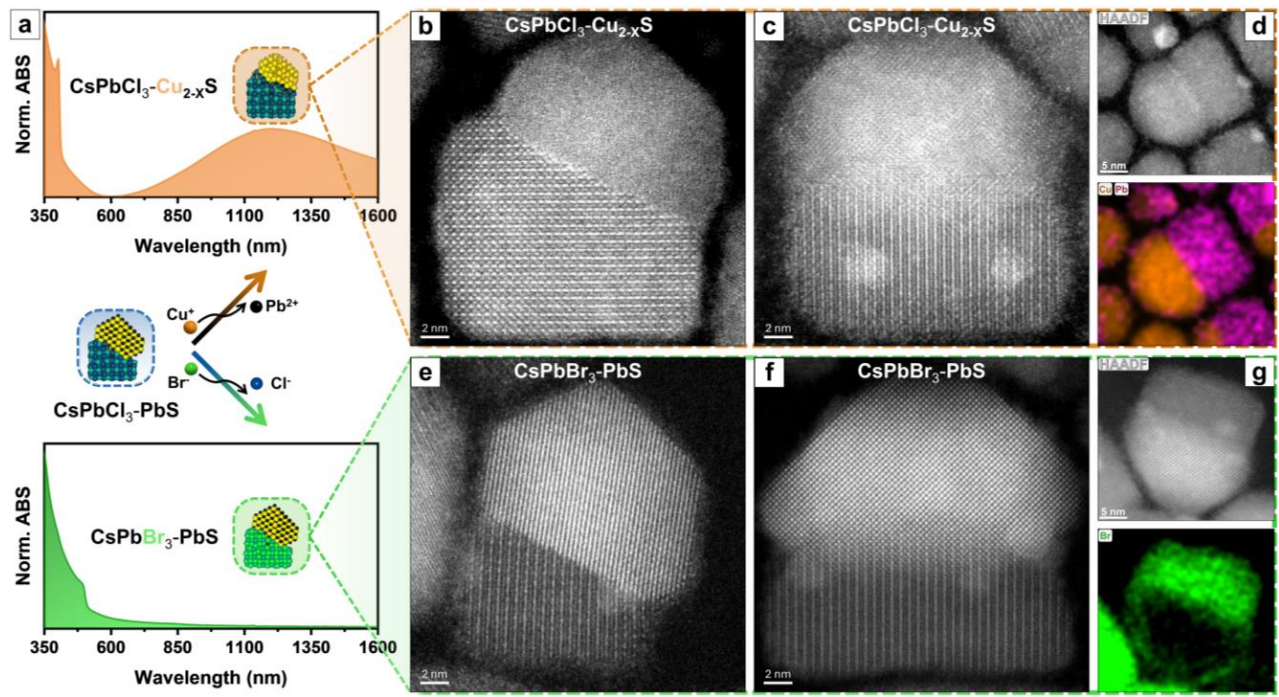

**Figure 4. Selective cation (Pb$^{2+}$→Cu$^+$) and anion (Cl$^-$→Br$^-$) exchange on CsPbCl$_3$/PbS heterostructure.** (a) Optical absorption spectra of cation (Pb$^{2+}$→Cu$^+$) exchanged CsPbCl$_3$/Cu$_{2-x}$S (upper), and anion (Cl$^-$→Br$^-$) exchanged CsPbBr$_3$/PbS heterostructures (down). (b-c) High-resolution HAADF STEM images of two different cations exchanged CsPbCl$_3$/Cu$_{2-x}$S heterostructures in common

orthogonal projections. (d) HAADF STEM image of a single CsPbCl$_3$/Cu$_{2-x}$S heterostructure with the corresponding STEM-EDX elemental map for Pb$^{2+}$ (purple), and Cu$^+$ (orange). (e-f) HAADF HR STEM- images of two different anion-exchanged CsPbBr$_3$/PbS heterostructures in orthogonal projections. (g) HAADF STEM image of a single CsPbBr$_3$/PbS heterostructure with the corresponding STEM-EDX elemental map for Br$^-$ (green).

The PLE spectrum (Figure S22) of the CsPbBr$_3$-PbS heterostructures closely resembles their absorption spectrum (Figure 5c). This similarity, observed in the parent heterostructures as well, suggests that photoexcited carriers in the CsPbBr$_3$ domain get transferred to the PbS domain and contribute to the NIR emission from PbS. Cation exchange of the same samples with Cu$^+$ revealed instead a broad absorption feature in the NIR spectral range, previously assigned to a localized surface plasmon resonance (Figure S20a)[51, 53] and attributed to the sub-stoichiometric composition of the digenite (Fm$\bar{3}$m) Cu$_{2-x}$S phase (Figure S19). Also, the Cu$^+$ treated heterostructure lost the IR emission originally stemming from the PbS domain. Such exchanges on the "220 °C/S-ODE" sample resulted in similar observations. For instance, the Br$^-$ exchange led to a red-shifted perovskite excitonic absorption (Figure 4a lower panel), whereas the Cu$^+$ treatment yielded a broad absorption in the NIR range (Figure 4a upper panel).

The distribution of the ions in the exchanged heterostructure samples was investigated by STEM-EDX. Compositional mapping of a single heterostructure treated by Cu$^+$ indicates pronounced segregation of Pb$^{2+}$ and Cu$^+$ ions within the two distinct domains of the heterostructure, CsPbCl$_3$, and Cu$_{2-x}$S, respectively (Figure 4d). This observation implies complete conversion from PbS to Cu$_{2-x}$S while retaining the integrity of the anionic framework, suggesting heterostructure with CsPbCl$_3$/Cu$_{2-x}$S composition. Additionally, elemental maps of a single CsPbBr$_3$/PbS heterostructure evidenced a homogeneous distribution of Br$^-$ ions throughout the perovskite domain (Figure 4g). EDX agrees with the XRD patterns and the optical measurements, confirming the full preservation of the PbS domain after the anion exchange. Even though neither Cs$^+$ nor Br$^-$ were present inside the PbS domain, both ions could be detected on its surface (Figure S23).

HAADF STEM analyses of the exchanged heterostructures also revealed similar epitaxial relationships with the parent CsPbCl$_3$-PbS heterostructure, since the FFT patterns demonstrate identical orientations with the constructed 3D model (Figure S23). This evidence highlights that the exchange did not disrupt the overall structure, proving that the anionic and cationic frameworks remained intact during the reactions. We identified the orientations and parallel planes for CsPbCl$_3$/Cu$_{2-x}$S through FFT analysis using two orthogonal projections of the heterostructure (Figure 4b,c). Figure 4b illustrates an atomic resolution image of the Cu$_{2-x}$S-CsPbCl$_3$ heterostructure in low index zone axes ([001] CsPbCl$_3$ and [1-10] Cu$_{2-x}$S). The corresponding FFTs show that the (-4-42) Cu$_{2-x}$S and (2-10) CsPbCl$_3$ planes are parallel and along the interface. These two sets of planes were also found to constitute the interface in the case of the parent PbS-CsPbCl$_3$ heterostructure. The FFT patterns (Figure S23f,g) indicate that the zone axes of the first projection (Figure 4b) are parallel planes in the second projection (Figure 4c), confirming the orthogonality of the two projections. A more careful analysis through STEM-EDX, HAADF STEM, and integrated Differential Phase Contrast (iDPC) STEM, as presented in Figures 5e,f, and S24 revealed the presence of a PbS monolayer between the CsPCl$_3$ and the Cu$_{2-x}$S domains. The resistance of this layer to the cation exchange reaction hints at comparatively higher stability, most likely because it is located at the interface (an aspect that will be discussed in more detail later). Data related to the CsPbBr$_3$/PbS heterostructure obtained by Cl$^-$→Br$^-$ exchange are presented in Figure 4e-g. HAADF STEM analyses of such cases unveiled a distinct interface between the two domains (Figure 4e,f), with the same epitaxial relationship observed in the CsPbCl$_3$/PbS case. This was further confirmed by the corresponding FFT patterns evidencing unexchanged orientations and planes of the interface (Figure S23).

The NIR emission of the exchanged CsPbBr$_3$/PbS heterostructure was redshifted with respect to the parent CsPbCl$_3$/PbS samples (Figure 5c). This red shift might point to a lower degree of confinement of the carriers in PbS for the CsPbBr$_3$/PbS heterostructures compared to the CsPbCl$_3$/PbS case. To further shed light on the heterostructures obtained by ion exchange reactions, we performed DFT calculations on CsPbBr$_3$/PbS and CsPbCl$_3$/Cu$_2$S NC models. For the CsPbBr$_3$/PbS heterostructure, we simply took the previously built CsPbCl$_3$/PbS model and systematically replaced the Cl$^-$ ions of the perovskite domain with Br$^-$ ions while preserving the cation sublattice. The relaxed structure of this model is depicted in Figure 5a. As in the chloride-based heterostructure, the stair-like epitaxial alignment of the CsPbBr$_3$/PbS NC ensures a six-fold coordination of the Pb$^{2+}$ cations at the interface between the perovskite and sulfide moieties. The electronic structure of this CsPbBr$_3$/PbS model, reported in Figure 5b, is closely aligned with that of the parent CsPbCl$_3$/PbS, with the main difference being a higher contribution of the Br-based perovskite to the valence band edge states, in line with a smaller band gap of CsPbbr$_3$ compared to CsPbCl$_3$. Overall, we see a slight blue shift of the bandgap of about 0.1 eV compared to CsPbCl$_3$/PbS, contrary to the experimentally observed red shift. However, the exact emission energy of the PbS moiety is difficult to discern due to the presence of interfacial states that introduce additional electronic levels within the bandgap, complicating the electronic structure landscape and hindering a more precise assignment of the emission characteristics of the PbS component.

To build the CsPbCl$_3$/Cu$_2$S heterostructure, we also started from the parent CsPbCl$_3$/PbS model and preserved the perovskite domain as well as one PbS layer at the interface (identified via HAADF STEM and iDPC STEM analysis), while replacing each Pb ion with two Cu ions in the remaining PbS domain. The latter operation was facilitated by the resemblance of the cubic bulk PbS and $Cu_2S$ crystallographic structure, which belong to the same $Fm\bar{3}m$ space group and share a common S anion sublattice. As anticipated and further evidenced in the relaxed structure of Figure 5d, the presence of an intermediate PbS layer results in a smooth transition between the two domains as it allows to complete of the coordination of the ions of both $CsPbCl_3$ and $Cu_2S$ domains at the interface. The electronic structure of the $CsPbCl_3/Cu_2S$ heterostructure, reported in Figure S25, exhibits no bandgap, mostly due to the presence of surface defects and an underestimation of the band gap typical of some of the DFT exchange-correlation functionals, like the one employed here.[54] We also point out that the model of $CsPbCl_3/Cu_2S$ heterostructures was prepared with a stoichiometric $Cu_2S$ composition, while the experiments revealed $Cu_{2-x}S$ based heterostructures, with the presence of $Cu^+$ vacancies, as seen from the absorption band in the NIR from that sample.

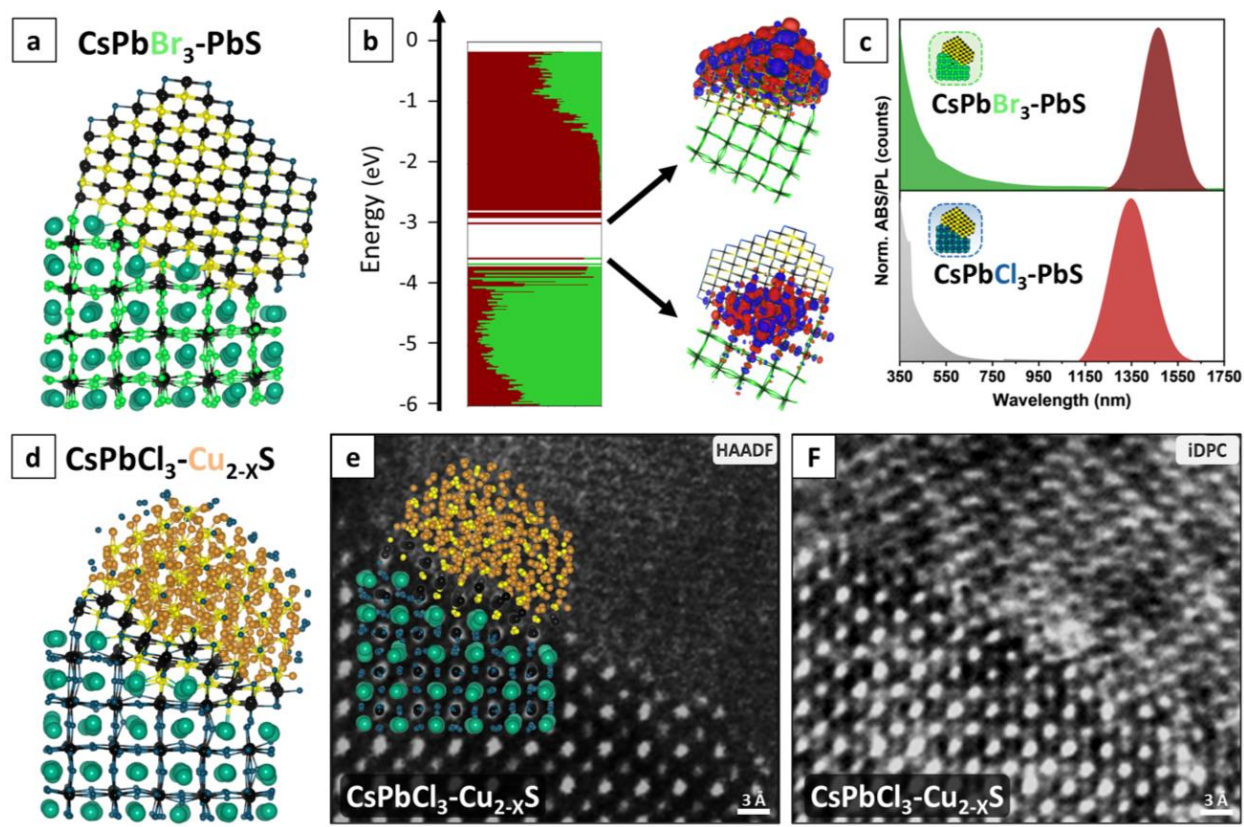

**Figure 5**. (a) Ball and sticks representation of the $CsPbBr_3/PbS$ NC heterostructure model optimized at the DFT/PBE level of theory. (b) The electronic structure of the model represented in (a) is computed at the DFT/PBE level of theory. The color code indicates the contribution of each domain to each molecular orbital. On the right, we plotted the frontier molecular orbitals to further highlight the resemblance with the parent $CsPbCl_3/PbS$ NC. (c) Optical absorption and PL spectra of pure $CsPbCl_3/PbS$ (down panel, sample "150 °C/TMS"), and the corresponding anion ($Cl^- \rightarrow Br^-$) exchanged $CsPbBr_3/PbS$ heterostructures (upper panel). (d) Ball and sticks representation of the $CsPbCl_3/Cu_2S$ NC heterostructure model optimized at the DFT/PBE level of theory. (e) High-resolution HAADF STEM image of cation exchanged $CsPbCl_3/Cu_{2-x}S$ highlighting the stair-like epitaxial interface. The superimposed DFT model confirms the presence of a PbS monolayer at the interface. (f) iDPC STEM image of the same area highlighting atomic columns of light elements (Cl, Cu, and S).

## CONCLUSIONS

We have reported the synthesis of epitaxial $CsPbCl_3/PbS$ heterostructures and analyzed their structural and optical features. The synthesis was made possible by introducing conditions that favor the nucleation and growth of PbS on top of the perovskite domain compared to the competing process of $Pb_4S_3Cl_2$ growth. This heterostructure was then used to generate additional heterostructures by selective anion/cation exchange on either the perovskite or the lead sulfide domain. While this is proven here just for $Cl^- \rightarrow Br^-$ and $Pb^{2+} \rightarrow Cu^+$ exchanges, the method should be extendable to encompass other reactions. This proof-of-concept study paves the way to a vast playground for potential new structures: those will likely include the replacement of the $Cs^+$ cation in the perovskite domain with other large monovalent cations such as methylammonium and formamidinium, and the $Pb^{2+}$ cation in the PbS domain with cations such as $Ag^+$, $Zn^{2+}$, $Cd^{2+}$, and many others, leading to heterostructures that might find applications in fields such as photocatalysis and photodetection.

**Experimental Section**

**Materials:** Cesium carbonate ($Cs_2CO_3$, 99,9%), lead chloride ($PbCl_2$, >98%), lead bromide ($PbBr_2$, >98%), lead acetate trihydrate ($Pb(CH_3COO)_2 \cdot 3H_2O$, 99.99%), manganese (II) acetate ($Mn(CH_3COO)_2$, 98%), sodium acetate ($C_2H_3NaO_2$ 99%), copper (I) oxide ($Cu_2O$, >99.9%), dodecanethiol (DDT, 99.9%), bis(trimethylsilyl) sulfide (($TMS)_2S$, $C_6H_{18}SSi_2$), elemental sulfur (S, >99%), 1-octadecene (ODE, $C_{18}H_{36}$, 90%) oleic acid (OA, $C_{18}H_{34}O_2$, 90%), oleylamine (OLAm, $C_{18}H_{37}N$, 70%), toluene ($C_7H_8$, >99.8%), ethyl acetate ($C_4H_8O_2$), acetone ($CH_3COCH_3$, >99.5%) dimethyl sulfoxide (($CH_3)_2SO$ >99.7%) and tetrachloroethane (TCE, $CHCl_2CHCl_2$, >98%), fumed silica powder ($SiO_2$), and hexane (99.8%) were purchased from Merck. All chemicals were used without further purification.

**Precursor solutions preparation**

**$PbCl_2$ stock solution.** $PbCl_2$ (2 mmol), 30 mL of octadecene, 5 mL of oleylamine, and 5 mL of oleic acid were loaded in a 100 mL 3-neck flask. The mixture was first degassed for 30 minutes at room temperature and then for 30 minutes at 110 °C under stirring. Finally, the solution was heated up to 150 °C under $N_2$ until the salt was completely dissolved. The resulting solution was transferred into a $N_2$-filled glass vial and was stored inside a glovebox for further use.

**Cs-OL stock solution.** $Cs_2CO_3$ (1 mmol), 2.5 mL of oleic acid, and 8.75 mL of octadecene were loaded in a 25 mL 3-neck flask. The mixture was first degassed for 30 minutes at room temperature and then for 1 hour at 110 °C under stirring. The resulting solution was transferred into a $N_2$-filled glass vial and was stored inside a glovebox for further use.

**Pb-OL stock solution**. 1 mmol of lead acetate trihydrate, 650 μL of oleic acid, and 9.35 mL of octadecene were loaded in a 50 mL 3-neck flask. The mixture was first degassed for 30 minutes at room temperature and then for 1 hour at 110 °C under stirring. The resulting solution was transferred into a $N_2$-filled glass vial and was stored inside a glovebox for further use.

**Mn-OL precursor solution.** 0.2 mmol of manganese acetate, 650 μL of oleic acid, and 9.35 mL of octadecene were loaded in a 50 mL 3-neck flask. The mixture was first degassed for 30 minutes at room temperature and then for 1 hour at 110 °C under stirring. The resulting solution was transferred into a $N_2$-filled glass vial and was stored inside a glovebox for further use.

**Na-OL precursor solution.** 0.1 mmol of sodium acetate, 650 μL of oleic acid, and 9.35 mL of octadecene were loaded in a 50 mL 3-neck flask. The mixture was first degassed for 30 minutes at room temperature and then for 1 hour at 110 °C under stirring. The resulting solution was transferred into a $N_2$-filled glass vial and was stored inside a glovebox for further use.

**S-ODE precursor solution.** 0.5 mmol of sulfur powder and 5 mL ODE (previously degassed for 1 hour at 110 °C) were loaded in a 7 mL vial inside a glovebox. The resulting mixture was sonicated until the sulfur powder was completely dissolved.

**TMS-ODE precursor solution.** 0.5 mmol of TMS and 5 mL ODE (previously degassed for 1 hour at 110 °C) were loaded and mixed in a 7 mL vial inside a glovebox.

**Cu-OL stock solution (cation exchange reactions).** 0.1 mmol $Cu_2O$, 650 μL of oleic acid, and 9.35 mL of octadecene were loaded in a 40 mL vial. The mixture was degassed first for 30 minutes at room temperature and then for 1 hour at 110 °C under stirring. The resulting solution was transferred into a $N_2$-filled glass vial and was stored inside a glovebox for further use.

**$PbBr_2$ stock solution (halide exchange reactions).** 1 mmol of $PbBr_2$, 2.5 mL of oleic acid, 2.5 mL of oleylamine, and 15 mL of octadecene were loaded in a 40 mL vial. The mixture was degassed for 30 minutes at room temperature and then for 30 minutes at 110 °C. Then, the solution was heated up to 150 °C under $N_2$ until the salt was completely dissolved. The resulting solution was transferred into a $N_2$-filled glass vial and was stored inside a glovebox for further use.

**Nanocrystal synthesis**

**$CsPbCl_3$ clusters synthesis.** $CsPbCl_3$ nanoclusters were synthesized following a previously reported method with slight modifications.[41] Briefly, 4 mL of the $PbCl_2$ stock solution was transferred into a 20 mL vial inside a glovebox. The solution was heated up to 50 °C and then 0.25 mL of Cs-OL stock solution was injected into the $PbCl_2$ stock solution. The resulting mixture was kept under stirring at 50 °C for approximately 30 min. After this time, the appearance of a white-turbid color indicated the assembly of $CsPbCl_3$ clusters formation. The resulting solution was purified by centrifugation (8000 rpm, 5 min), and the supernatant was discarded. Finally, the precipitate was redispersed in a solution containing 0.9 mL of previously degassed octadecene and 0.3 mL of Mn-OL (or Na-OL) solution.

**$CsPbCl_3$-PbS heterostructures synthesis.** In a typical synthesis, 4.0 mL of previously degassed octadecene was loaded into a 20 mL $N_2$-filled vial. Then, the vial was heated up to the corresponding reaction temperature (125 - 220 °C) for 5 min. After this time, the following precursor solutions were swiftly injected into the octadecene: 0.1 mL of Pb-OL solution, 0.2 mL of DDT-ODE solution, clusters solution (mentioned above), and 0.1 mL of the sulfur source (S-ODE for large $CsPbCl_3$-PbS heterostructures ("220 °C/S-ODE" synthesis) or TMS-ODE for small and pure $CsPbCl_3$-PbS heterostructures ("150 °C/TMS" synthesis)). The reaction was kept under stirring for the corresponding reaction time (3-20 min) and finally quenched in an ice-water bath. The heterostructures were purified by centrifugation (6000 rpm, 5 min), the supernatant was discarded, and the precipitate was redispersed in toluene.

For concentration reference, the absorption spectrum of the NCs solution was measured (after a dilution of 50 μL NCs to 2.5 mL toluene) presenting an optical density of 0.26 at 370 nm.

**Exchange reactions**

**Cu-exchange reactions.** Cation exchange reactions were performed in a glovebox. In a typical reaction, 750 μL of the $CsPbCl_3$-PbS solution and 500 μL of Cu-OL stock solution were loaded in a 7 mL vial. The reaction was kept overnight under stirring at room temperature. Afterwards, the heterostructures were purified with EtOAc (600 μL) and collected by centrifugation (6000 rpm, 5 min). The supernatant was discarded, and the precipitate was redispersed in toluene for further characterization.

**Br-exchange reactions.** Halide exchange reactions were performed in a glovebox. In a typical reaction, 750 μL of the $CsPbCl_3$-PbS solution and 500 μL of $PbBr_2$ stock solution were loaded in a 7 mL vial. The reaction was kept overnight under stirring at room temperature. Then, the heterostructures were purified with EtOAc (600 μL) and collected by centrifugation (6000 rpm, 5 min). The supernatant was discarded, and the precipitate was redispersed in toluene for further characterization.

**$CsPbCl_3$-PbS heterostructures etching.** PbS NCs were obtained from the etching of $CsPbCl_3$-PbS heterostructures. The $CsPbCl_3$ domain was etched from the heterostructures by dissolving it in a mixture of $CsPbCl_3$-PbS (1 mL), DMSO (0.25 mL), and acetone (0.25 mL). The resulting PbS NCs were collected by centrifugation (6000 rpm, 5 min) and redispersed in toluene.

**Optical characterization.** UV-Vis absorption spectra were carried out using a Varian Cary 5000 UV-Vis-NIR absorption spectrophotometer (Agilent). The spectra were collected by diluting 50 μL of the sample in toluene in 2.5 mL of TCE. UV-Vis photoluminescence spectra were obtained on a Varian Cary Eclipse Spectrophotometer (Agilent) using $\lambda_{ex}$= 370 nm. IR photoluminescence spectra were obtained on an Edinburgh FS5 spectrofluorometer using $\lambda_{ex}$= 370 nm. UV-Vis-NIR excitation spectra were obtained on an Edinburgh FS5 spectrofluorometer. The experimental emission spectra were fitted (Gauss fit) to remove noise in the data in the NIR spectral region.

**Transmission Electron Microscopy.** Bright-field TEM images of the samples were acquired with a JEOL-1400Plus transmission electron microscope operating at an acceleration voltage of 100 kV and on a JEOL JEM 1011 transmission electron microscope operating at an acceleration voltage of 120 kV. Samples were prepared by drop casting from diluted NC solutions onto carbon film-coated 200 mesh copper grids.

**X-ray Powder Diffraction.** X-ray powder diffraction measurements were performed on a PANanalytical Empyrean X-ray diffractometer, equipped with a 1.8 kW Cu Kα ceramic anode and a PIXcel3D 2 × 2 area detector, operating at 45 kV and 40 mA. NCs' dispersions were mixed with fumed silica and dried to minimize the preferential orientation. Finally, the powder was placed on a zero-diffraction silicon substrate to perform the measurements.

**(Scanning) Transmission electron microscopy ((S)TEM) characterization.** High-resolution scanning TEM (HRSTEM) images were acquired on a probe-corrected ThermoFisher Spectra 30-300 S/TEM operated at 300 kV, using a HAADF detector with a beam current of a few tens of pA to limit beam damage to the s. The iDPC images were collected on a segmented Panther detector; iDPC, compared to HAADF, provides more contrast for light elements. The convergence angle was set to 25 mrad, corresponding to a sub-angstrom electron beam. Compositional maps were acquired using Velox, with a probe current of ~150 pA and rapid rastered scanning. The Energy-Dispersive X-Ray (EDX) spectroscopy signal was acquired on a Dual-X setup comprising two detectors on either side of the sample, for a total acquisition solid angle of 1.76 Sr.

**Computational methodology.** To shed light onto the structural and electronic properties of the $PbS/CsPbCl_3$, $PbS/CsPbBr_3$, and $Cu_2S/CsPbCl_3$ heterostructured NCs, we have carried out atomistic simulations at the density functional theory (DFT) level using the PBE exchange–correlation functional[55] and a double-ζ basis set plus polarization functions on all atoms[56] as implemented in the CP2K 6.1 package.[57] Scalar relativistic effects were incorporated as effective core potential functions in the basis set.[56] All structures have been optimized in vacuum. Details on how the models were prepared are reported in the main text. For the parent $PbS/CsPbCl_3$ we have additionally computed the inverse participation ratio (IPR)[58] in order to identify surface localized states. As demonstrated for other NCs,[48] the IPR quantifies the orbital localization of a given molecular orbital and is defined as:

$$IPR_i = \frac{\sum_\alpha |P_{\alpha,i}|^4}{(\sum_\alpha |P_{\alpha,i}|^2)^2}$$

Here, $P_{\alpha,i}$ represents the weight of molecular orbital $i$ on a given atom $\alpha$ expanded on an atomic orbital basis. For finite systems, the IPR provides an estimate of the number of atoms that contribute to a given electronic state $i$. It can range from the inverse of the number of atoms in the system when the wave function is distributed equally over all atoms in the system to 1 in the case of a localized state to a single atom. When values are near zero, the IPR identifies delocalized states.

To verify that the presence $Mn^{2+}$ ions could modify the fraction of the available $Cl^-$ ions in solution, we also computed the binding affinity of $MnCl_2$ and $PbCl_2$, defined as:

$$E_{binding} = E_{MCl_2} - (E_{M^{2+}} + E_{Cl^-})$$

where the $MCl_2$ molecular complexes structures have been relaxed in vacuum at the DFT/PBE/DZVP level of theory. The $Mn^{2+}$ ions have been considered in an open-shell sextuplet electron configuration.

## ASSOCIATED CONTENT

**Supporting Information**. Additional details on control syntheses, stability tests, additional data on optical, structural, morphological and compositional characterization. This material is available free of charge via the Internet at http://pubs.acs.org.

## AUTHOR INFORMATION

### Corresponding Author

* Ivan Infante (ivan.infante@bcmaterials.net>)
* Giorgio Divitini (giorgio.divitini@iit.it)
* Liberato Manna (liberato.manna@iit.it)

### Author Contributions

The manuscript was written through the contributions of all authors.

### Funding Sources

J.Z. and L.M. acknowledge funding from the Project IEMAP (Italian Energy Materials Acceleration Platform) within the Italian Research Program ENEA-MASE (Ministero dell'Ambiente e della Sicurezza Energetica) 2021-2024 "Mission Innovation" (agreement 21A033302 GU n. 133/5-6-2021). L.M. acknowledges funding from European Research Council through the ERC Advanced Grant NEHA (grant agreement n. 101095974). G.D. and L.M. acknowledge funding from the Italian Space Agency - contract ASI N. 2023-4-U.0.

## ACKNOWLEDGMENT

We acknowledge the materials characterization facility at Istituto Italiano di Tecnologia providing access to the PANalytical Empyrean X-Ray Diffractometer. We also acknowledge the computing resources and the related technical support used for this work have been provided by CRESCO/ENEAGRID High-Performance Computing infrastructure and its staff.[59] CRESCO/ENEAGRID High-Performance Computing infrastructure is funded by ENEA, the Italian National Agency for New Technologies, Energy and Sustainable Economic Development, and by Italian and European research programs (http://www.cresco.enea.it/english)

# REFERENCES

(1) Protesescu, L.; Yakunin, S.; Bodnarchuk, M. I.; Krieg, F.; Caputo, R.; Hendon, C. H.; Yang, R. X.; Walsh, A.; Kovalenko, M. V. Nanocrystals of Cesium Lead Halide Perovskites ($CsPbX_3$, X = Cl, Br, and I): Novel Optoelectronic Materials Showing Bright Emission with Wide Color Gamut. *Nano Lett.* **2015**, *15* (6), 3692-3696.

(2) Guzelturk, B.; Winkler, T.; Van de Goor, T. W. J.; Smith, M. D.; Bourelle, S. A.; Feldmann, S.; Trigo, M.; Teitelbaum, S. W.; Steinrück, H.-G.; de la Pena, G. A.; et al. Visualization of dynamic polaronic strain fields in hybrid lead halide perovskites. *Nat. Mater.* **2021**, *20* (5), 618-623.

(3) Maes, J.; Balcaen, L.; Drijvers, E.; Zhao, Q.; De Roo, J.; Vantomme, A.; Vanhaecke, F.; Geiregat, P.; Hens, Z. Light Absorption Coefficient of $CsPbBr_3$ Perovskite Nanocrystals. *J. Phys. Chem. Lett.* **2018**, *9* (11), 3093-3097.

(4) Di Stasio, F.; Christodoulou, S.; Huo, N.; Konstantatos, G. Near-Unity Photoluminescence Quantum Yield in $CsPbBr_3$ Nanocrystal Solid-State Films via Postsynthesis Treatment with Lead Bromide. *Chem. Mater.* **2017**, *29* (18), 7663-7667.

(5) Kovalenko, M. V.; Protesescu, L.; Bodnarchuk, M. I. Properties and potential optoelectronic applications of lead halide perovskite nanocrystals. *Science* **2017**, *358* (6364), 745-750.

(6) Yang, Y.; Wang, D.; Li, Y.; Xia, J.; Wei, H.; Ding, C.; Hu, Y.; Wei, Y.; Li, H.; Liu, D.; et al. In Situ Room-Temperature Synthesis of All-Colloidal Quantum Dot $CsPbBr_3$–PbS Heterostructures. *ACS Photonics* **2023**, *10* (12), 4305-4314.

(7) Bera, S.; Pradhan, N. Perovskite Nanocrystal Heterostructures: Synthesis, Optical Properties, and Applications. *ACS Energy Lett.* **2020**, *5* (9), 2858-2872.

(8) Jia, C.; Li, H.; Meng, X.; Li, H. $CsPbX_3/Cs_4PbX_6$ core/shell perovskite nanocrystals. *Chem. Commun.* **2018**, *54* (49), 6300-6303.

(9) Toso, S.; Imran, M.; Mugnaioli, E.; Moliterni, A.; Caliandro, R.; Schrenker, N. J.; Pianetti, A.; Zito, J.; Zaccaria, F.; Wu, Y.; et al. Halide perovskites as disposable epitaxial templates for the phase-selective synthesis of lead sulfochloride nanocrystals. *Nat. Commun.* **2022**, *13* (1), 3976.

(10) Rusch, P.; Toso, S.; Ivanov, Y. P.; Marras, S.; Divitini, G.; Manna, L. Nanocrystal Heterostructures Based On Halide Perovskites and Lead–Bismuth Chalcogenides. *Chem. Mater.* **2023**, *35* (24), 10684-10693.

(11) Qiu, H.; Li, F.; He, S.; Shi, R.; Han, Y.; Abudukeremu, H.; Zhang, L.; Zhang, Y.; Wang, S.; Liu, W.; et al. Epitaxial $CsPbBr_3$/CdS Janus Nanocrystal Heterostructures for Efficient Charge Separation. *Adv. Sci.* **2023**, *10* (13), 2206560.

(12) Das, R.; Patra, A.; Dutta, S. K.; Shyamal, S.; Pradhan, N. Facets-Directed Epitaxially Grown Lead Halide Perovskite-Sulfobromide Nanocrystal Heterostructures and Their Improved Photocatalytic Activity. *J. Am. Chem. Soc.* **2022**, *144* (40), 18629-18641.

(13) Zhang, X.; Wu, X.; Liu, X.; Chen, G.; Wang, Y.; Bao, J.; Xu, X.; Liu, X.; Zhang, Q.; Yu, K.; et al. Heterostructural $CsPbX_3$-PbS (X = Cl, Br, I) Quantum Dots with Tunable Vis–NIR Dual Emission. *J. Am. Chem. Soc.* **2020**, *142* (9), 4464-4471.

(14) Kipkorir, A.; DuBose, J.; Cho, J.; Kamat, P. V. $CsPbBr_3$–CdS heterostructure: stabilizing perovskite nanocrystals for photocatalysis. *Chem. Sci.* **2021**, *12* (44), 14815-14825.

(15) Shamsi, J.; Dang, Z.; Ijaz, P.; Abdelhady, A. L.; Bertoni, G.; Moreels, I.; Manna, L. Colloidal CsX (X = Cl, Br, I) Nanocrystals and Their Transformation to $CsPbX_3$ Nanocrystals by Cation Exchange. *Chem. Mater.* **2018**, *30* (1), 79-83.

(16) Baranov, D.; Caputo, G.; Goldoni, L.; Dang, Z.; Scarfiello, R.; De Trizio, L.; Portone, A.; Fabbri, F.; Camposeo, A.; Pisignano, D.; et al. Transforming colloidal $Cs_4PbBr_6$ nanocrystals with poly(maleic anhydride-alt-1-octadecene) into stable $CsPbBr_3$ perovskite emitters through intermediate heterostructures. *Chem. Sci.* **2020**, *11* (15), 3986-3995.

(17) Patra, A.; Jagadish, K.; Ravishankar, N.; Pradhan, N. Epitaxial Heterostructures of $CsPbBr_3$ Perovskite Nanocrystals with Post-transition Metal Bismuth. *Nano Lett.* **2024**, *24* (5), 1710-1716.

(18) Behera, R. K.; Jagadish, K.; Shyamal, S.; Pradhan, N. Pt-$CsPbBr_3$ Perovskite Nanocrystal Heterostructures: All Epitaxial. *Nano Lett.* **2023**, *23* (17), 8050-8056.

(19) Rodríguez Ortiz, F. A.; Roman, B. J.; Wen, J.-R.; Mireles Villegas, N.; Dacres, D. F.; Sheldon, M. T. The role of gold oxidation state in the synthesis of Au-$CsPbX_3$ heterostructure or lead-free $Cs_2Au^IAu^{III}X_6$ perovskite nanoparticles. *Nanoscale* **2019**, *11* (39), 18109-18115.

(20) Balakrishnan, S. K.; Kamat, P. V. Au–$CsPbBr_3$ Hybrid Architecture: Anchoring Gold Nanoparticles on Cubic Perovskite Nanocrystals. *ACS Energy Lett.* **2017**, *2* (1), 88-93.

(21) Qiao, B.; Song, P.; Cao, J.; Zhao, S.; Shen, Z.; Di, G.; Liang, Z.; Xu, Z.; Song, D.; Xu, X. Water-resistant, monodispersed and stably luminescent $CsPbBr_3$/$CsPb_2Br_5$ core–shell-like structure lead halide perovskite nanocrystals. *Nanotechnol.* **2017**, *28* (44), 445602.

(22) Dutta, S. K.; Bera, S.; Pradhan, N. Why Is Making Epitaxially Grown All Inorganic Perovskite–Chalcogenide Nanocrystal Heterostructures Challenging? Some Facts and Some Strategies. *Chem. Mater.* **2021**, *33* (11), 3868-3877.

(23) Imran, M.; Peng, L.; Pianetti, A.; Pinchetti, V.; Ramade, J.; Zito, J.; Di Stasio, F.; Buha, J.; Toso, S.; Song, J.; et al. Halide Perovskite–Lead Chalcohalide Nanocrystal Heterostructures. *J. Am. Chem. Soc.* **2021**, *143* (3), 1435-1446.

(24) Smith, A. M.; Nie, S. Semiconductor Nanocrystals: Structure, Properties, and Band Gap Engineering. *Acc. Chem. Res.* **2010**, *43* (2), 190-200.

(25) Müller, J.; Lupton, J. M.; Lagoudakis, P. G.; Schindler, F.; Koeppe, R.; Rogach, A. L.; Feldmann, J.; Talapin, D. V.; Weller, H. Wave Function Engineering in Elongated Semiconductor Nanocrystals with Heterogeneous Carrier Confinement. *Nano Lett.* **2005**, *5* (10), 2044-2049.

(26) Ivanov, S. A.; Nanda, J.; Piryatinski, A.; Achermann, M.; Balet, L. P.; Bezel, I. V.; Anikeeva, P. O.; Tretiak, S.; Klimov, V. I. Light amplification using inverted core/shell nanocrystals: towards lasing in the single-exciton regime. *J. Phys. Chem. B* **2004**, *108* (30), 10625-10630.

(27) Kim, S.; Fisher, B.; Eisler, H.-J.; Bawendi, M. Type-II quantum dots: CdTe/CdSe (core/shell) and CdSe/ZnTe (core/shell) heterostructures. *J. Am. Chem. Soc.* **2003**, *125* (38), 11466-11467.

(28) Pal, B. N.; Ghosh, Y.; Brovelli, S.; Laocharoensuk, R.; Klimov, V. I.; Hollingsworth, J. A.; Htoon, H. 'Giant'CdSe/CdS core/shell nanocrystal quantum dots as efficient electroluminescent materials: strong influence of shell thickness on light-emitting diode performance. *Nano Lett.* **2012**, *12* (1), 331-336.

(29) Sagar, L. K.; Bappi, G.; Johnston, A.; Chen, B.; Todorović, P.; Levina, L.; Saidaminov, M. I.; García de Arquer, F. P.; Hoogland, S.; Sargent, E. H. Single-precursor intermediate shelling enables bright, narrow line width InAs/InZnP-based QD emitters. *Chem. Mater.* **2020**, *32* (7), 2919-2925.
(30) Zhang, X.; Fu, Q.; Duan, H.; Song, J.; Yang, H. Janus nanoparticles: From fabrication to (bio) applications. *ACS Nano* **2021**, *15* (4), 6147-6191.
(31) Reiss, P.; Bleuse, J.; Pron, A. Highly luminescent CdSe/ZnSe core/shell nanocrystals of low size dispersion. *Nano Lett.* **2002**, *2* (7), 781-784.
(32) Vaneski, A.; Schneider, J.; Susha, A. S.; Rogach, A. L. Colloidal hybrid heterostructures based on II–VI semiconductor nanocrystals for photocatalytic hydrogen generation. *J. Photochem. Photobiol. C: Photochem. Rev.* **2014**, *19*, 52-61.
(33) An, M. N.; Park, S.; Brescia, R.; Lutfullin, M.; Sinatra, L.; Bakr, O. M.; De Trizio, L.; Manna, L. Low-temperature molten salts synthesis: $CsPbBr_3$ nanocrystals with high photoluminescence emission buried in mesoporous $SiO_2$. *ACS Energy Lett.* **2021**, *6* (3), 900-907.
(34) Zhong, Q.; Cao, M.; Hu, H.; Yang, D.; Chen, M.; Li, P.; Wu, L.; Zhang, Q. One-Pot Synthesis of Highly Stable $CsPbBr_3@SiO_2$ Core–Shell Nanoparticles. *ACS Nano* **2018**, *12* (8), 8579-8587.
(35) Liu, H.; Tan, Y.; Cao, M.; Hu, H.; Wu, L.; Yu, X.; Wang, L.; Sun, B.; Zhang, Q. Fabricating $CsPbX_3$-based type I and type II heterostructures by tuning the halide composition of janus $CsPbX_3/ZrO_2$ nanocrystals. *ACS Nano* **2019**, *13* (5), 5366-5374.
(36) Li, Z. J.; Hofman, E.; Li, J.; Davis, A. H.; Tung, C. H.; Wu, L. Z.; Zheng, W. Photoelectrochemically active and environmentally stable $CsPbBr_3/TiO_2$ core/shell nanocrystals. *Adv. Funct. Mater.* **2018**, *28* (1), 1704288.
(37) Chen, W.; Hao, J.; Hu, W.; Zang, Z.; Tang, X.; Fang, L.; Niu, T.; Zhou, M. Enhanced stability and tunable photoluminescence in perovskite $CsPbX_3/ZnS$ quantum dot heterostructure. *Small* **2017**, *13* (21), 1604085.
(38) Ravi, V. K.; Saikia, S.; Yadav, S.; Nawale, V. V.; Nag, A. $CsPbBr_3/ZnS$ Core/Shell Type Nanocrystals for Enhancing Luminescence Lifetime and Water Stability. *ACS Energy Lett.* **2020**, *5* (6), 1794-1796.
(39) Jagadeeswararao, M.; Vashishtha, P.; Hooper, T. J. N.; Kanwat, A.; Lim, J. W. M.; Vishwanath, S. K.; Yantara, N.; Park, T.; Sum, T. C.; Chung, D. S.; et al. One-Pot Synthesis and Structural Evolution of Colloidal Cesium Lead Halide–Lead Sulfide Heterostructure Nanocrystals for Optoelectronic Applications. *J. Phys. Chem. Lett.* **2021**, *12* (39), 9569-9578.
(40) Hassan, M. S.; Basera, P.; Bera, S.; Mittal, M.; Ray, S. K.; Bhattacharya, S.; Sapra, S. Enhanced photocurrent owing to shuttling of charge carriers across 4-aminothiophenol-functionalized $MoSe_2$–$CsPbBr_3$ nanohybrids. *ACS Appl. Mater. Interfaces* **2020**, *12* (6), 7317-7325.
(41) Peng, L.; Dutta, A.; Xie, R.; Yang, W.; Pradhan, N. Dot–wire–platelet–cube: step growth and structural transformations in $CsPbBr_3$ perovskite nanocrystals. *ACS Energy Lett.* **2018**, *3* (8), 2014-2020.
(42) *Lange's Handbook of Chemistry*; McGraw-Hill Education, 2017.
(43) Ning, Z.; Gong, X.; Comin, R.; Walters, G.; Fan, F.; Voznyy, O.; Yassitepe, E.; Buin, A.; Hoogland, S.; Sargent, E. H. Quantum-dot-in-perovskite solids. *Nature* **2015**, *523* (7560), 324-328.
(44) Toso, S.; Akkerman, Q. A.; Martín-García, B.; Prato, M.; Zito, J.; Infante, I.; Dang, Z.; Moliterni, A.; Giannini, C.; Bladt, E. Nanocrystals of lead chalcohalides: a series of kinetically trapped metastable nanostructures. *J. Am. Chem. Soc.* **2020**, *142* (22), 10198-10211.
(45) Wells, R. L.; Pitt, C. G.; McPhail, A. T.; Purdy, A. P.; Shafieezad, S.; Hallock, R. B. The use of tris (trimethylsilyl) arsine to prepare gallium arsenide and indium arsenide. *Chem. Mater.* **1989**, *1* (1), 4-6.
(46) ten Brinck, S.; Infante, I. Surface termination, morphology, and bright photoluminescence of cesium lead halide perovskite nanocrystals. *ACS Energy Lett.* **2016**, *1* (6), 1266-1272.
(47) Bodnarchuk, M. I.; Boehme, S. C.; Ten Brinck, S.; Bernasconi, C.; Shynkarenko, Y.; Krieg, F.; Widmer, R.; Aeschlimann, B.; Günther, D.; Kovalenko, M. V. Rationalizing and controlling the surface structure and electronic passivation of cesium lead halide nanocrystals. *ACS Energy Lett.* **2018**, *4* (1), 63-74.
(48) Houtepen, A. J.; Hens, Z.; Owen, J. S.; Infante, I. On the origin of surface traps in colloidal II–VI semiconductor nanocrystals. *Chem. Mater.* **2017**, *29* (2), 752-761.
(49) Akkerman, Q. A.; Genovese, A.; George, C.; Prato, M.; Moreels, I.; Casu, A.; Marras, S.; Curcio, A.; Scarpellini, A.; Pellegrino, T. From binary $Cu_2S$ to ternary Cu–In–S and quaternary Cu–In–Zn–S nanocrystals with tunable composition via partial cation exchange. *ACS Nano* **2015**, *9* (1), 521-531.
(50) De Trizio, L.; Li, H.; Casu, A.; Genovese, A.; Sathya, A.; Messina, G. C.; Manna, L. Sn cation valency dependence in cation exchange reactions involving $Cu_{2-x}Se$ nanocrystals. *J. Am. Chem. Soc.* **2014**, *136* (46), 16277-16284.
(51) Xie, Y.; Riedinger, A.; Prato, M.; Casu, A.; Genovese, A.; Guardia, P.; Sottini, S.; Sangregorio, C.; Miszta, K.; Ghosh, S. Copper sulfide nanocrystals with tunable composition by reduction of covellite nanocrystals with $Cu^+$ ions. *J. Am. Chem. Soc.* **2013**, *135* (46), 17630-17637.
(52) De Trizio, L.; Manna, L. Forging colloidal nanostructures via cation exchange reactions. *J. Chem. Rev.* **2016**, *116* (18), 10852-10887.
(53) Xie, Y.; Carbone, L.; Nobile, C.; Grillo, V.; D'Agostino, S.; Della Sala, F.; Giannini, C.; Altamura, D.; Oelsner, C.; Kryschi, C. Metallic-like stoichiometric copper sulfide nanocrystals: phase-and shape-selective synthesis, near-infrared surface plasmon resonance properties, and their modeling. *ACS Nano* **2013**, *7* (8), 7352-7369.
(54) Van Meer, R.; Gritsenko, O.; Baerends, E. Physical meaning of virtual Kohn–Sham orbitals and orbital energies: an ideal basis for the description of molecular excitations. *J. Chem. Theory Comput.* **2014**, *10* (10), 4432-4441.
(55) Perdew, J. P.; Burke, K.; Ernzerhof, M. Generalized gradient approximation made simple. *Phys. Rev. Lett.* **1996**, *77* (18), 3865.
(56) VandeVondele, J.; Hutter, J. Gaussian basis sets for accurate calculations on molecular systems in gas and condensed phases. *J. Chem. Phys.* **2007**, *127* (11).
(57) Hutter, J.; Iannuzzi, M.; Schiffmann, F.; VandeVondele, J. cp2k: atomistic simulations of condensed matter systems. *Wiley Interdiscip. Rev. Comput. Mol. Sci.* **2014**, *4* (1), 15-25.
(58) Murphy, N. C.; Wortis, R.; Atkinson, W. A. Generalized inverse participation ratio as a possible measure of localization for interacting systems. *Phys. Rev. B* **2011**, *83* (18), 184206.
(59) Iannone, F.; Ambrosino, F.; Bracco, G.; De Rosa, M.; Funel, A.; Guarnieri, G.; Migliori, S.; Palombi, F.; Ponti, G.; Santomauro, G. CRESCO ENEA HPC clusters: a working example of a multifabric GPFS Spectrum Scale layout. In *2019 International Conference on High Performance Computing & Simulation (HPCS)*, 2019; IEEE: pp 1051-1052.

TOC:

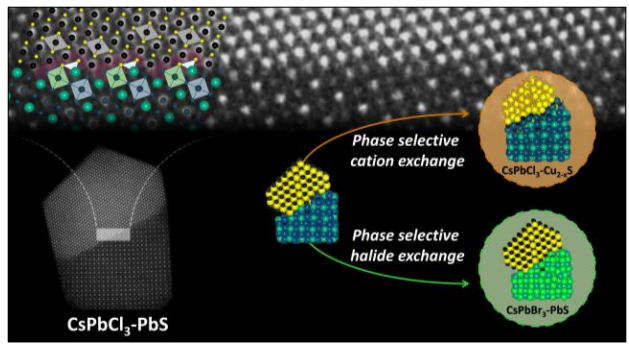

Supporting Information for

# Nanocrystals Heterostructures based on Halide Perovskites and Metal Sulfides

Nikolaos Livakas[1,2], Juliette Zito[1], Yurii P. Ivanov[3], Clara Otero Martínez[4], Giorgio Divitini[3*], Ivan Infante[5,6*], Liberato Manna[1*]

[1] Nanochemistry, Istituto Italiano di Tecnologia, Via Morego 30, Genova, Italy

[2] Dipartimento di Chimica e Chimica Industriale, Università di Genova, 16146 Genova, Italy

[3] Electron Spectroscopy and Nanoscopy, Istituto Italiano di Tecnologia, Via Morego 30, Genova, Italy

[4] CINBIO, Department of Physical Chemistry, Materials Chemistry and Physics Group, Universidade de Vigo, Campus Universitario As Lagoas-Marcosende, 36310 Vigo, Spain

[5] BCMaterials, Basque Center for Materials, Applications, and Nanostructures, UPV/EHU Science Park, Leioa 48940, Spain

[6] Ikerbasque Basque Foundation for Science Bilbao 48009, Spain

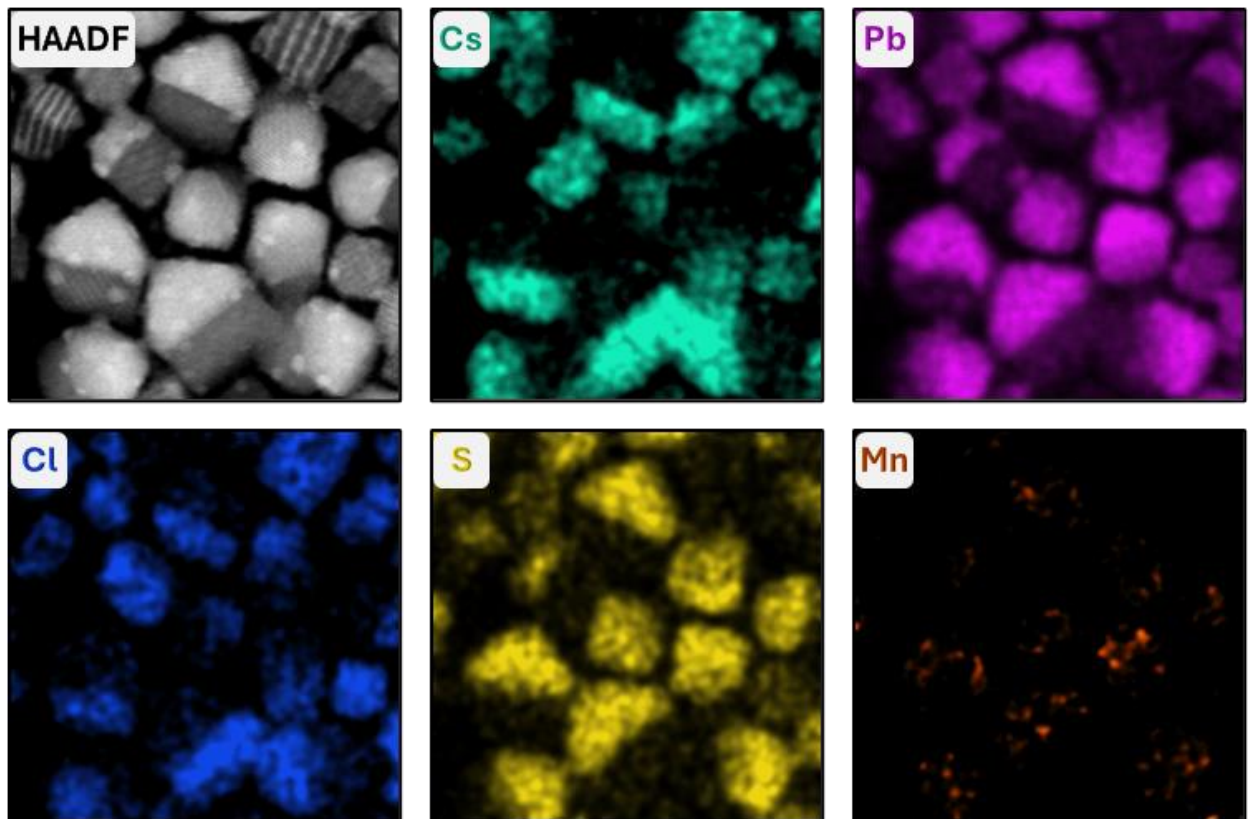

**Figure S1.** STEM-HAADF image of CsPbCl$_3$-PbS heterostructures (case sample "220 °C/S-ODE", presented in Figure 1a) with the corresponding EDX elemental maps for cesium, lead, chlorine, sulfur, and manganese indicating a small presence of manganese on the nanocrystal surface.

**Table S1.** Summary of the elemental analysis obtained from the EDX spectrum of **Figure S1** expressed in atomic percentage (%) and mass percentage (%).

| Z | Element | Atomic (%) | Mass (%) | Fit Error (%) |
|---|---------|------------|----------|---------------|
| 16 | S | 21.42 | 5.32 | 3.79 |
| 17 | Cl | 17.18 | 4.72 | 0.73 |
| 25 | Mn | 1.74 | 0.74 | 1.06 |
| 55 | Cs | 11.43 | 11.77 | 0.05 |
| 82 | Pb | 48.23 | 77.45 | 0.09 |

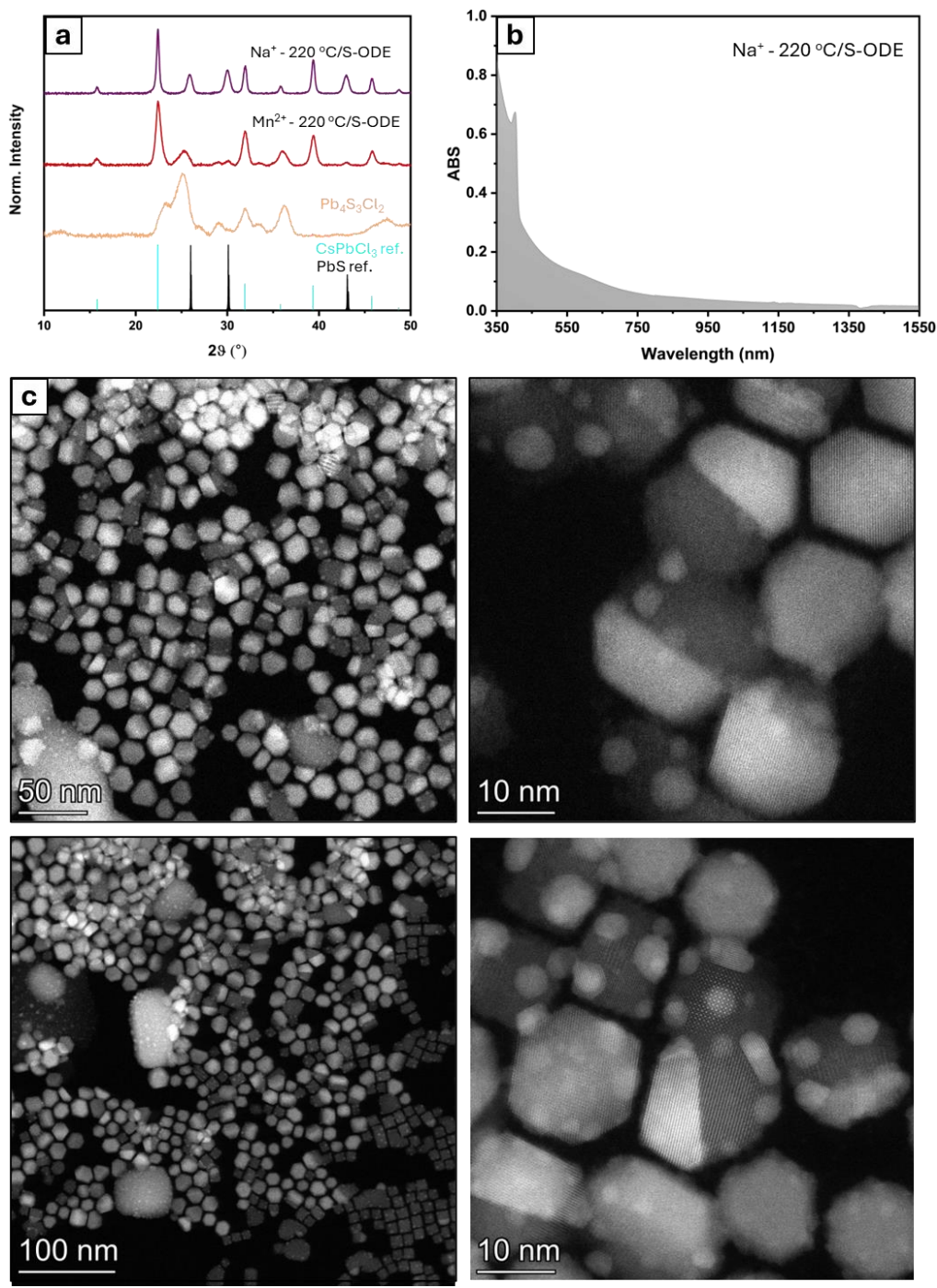

**Figure S2.** (a) XRD diffraction patterns of the synthesis route "220 °C/S-ODE" in which either Na-OL (purple) or Mn-OL (red) is added. We report, for comparison, the diffraction pattern of a sample of $Pb_4S_3Cl_2$ chalcohalide nanocrystals (orange) and $CsPbCl_3$, and PbS bulk reference patterns (cyan and black, respectively). The peak at ~ 43° indicates the formation PbS. (b) Optical absorption spectrum of the sample from the synthesis "220 °C/S-ODE" in which Na-OL has been used instead of Mn-OL. (c) STEM-HAADF images of the same sample as in (b).

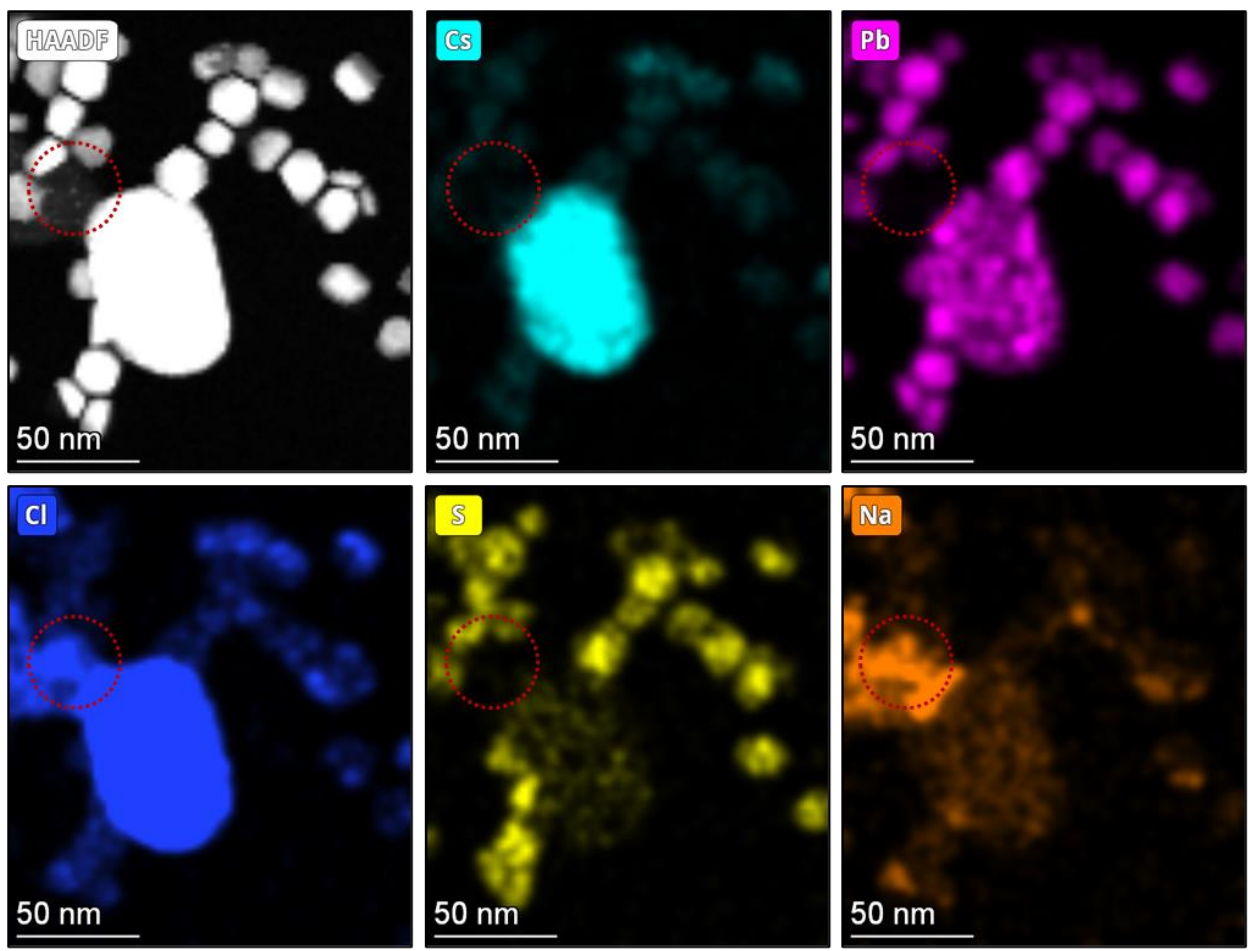

**Figure S3.** STEM-HAADF images of the sample case "220 °C/S-ODE" in which Na-OL has been used instead of Mn-OL, along with the corresponding EDX elemental maps for cesium, lead, chlorine, sulfur, and sodium. The area marked with a red dashed cycle corresponds to an amorphous aggregate containing mainly Na and Cl. A large nanocrystal, containing mainly Cs, Pb, and Cl (along with traces of S and Na) is also present.

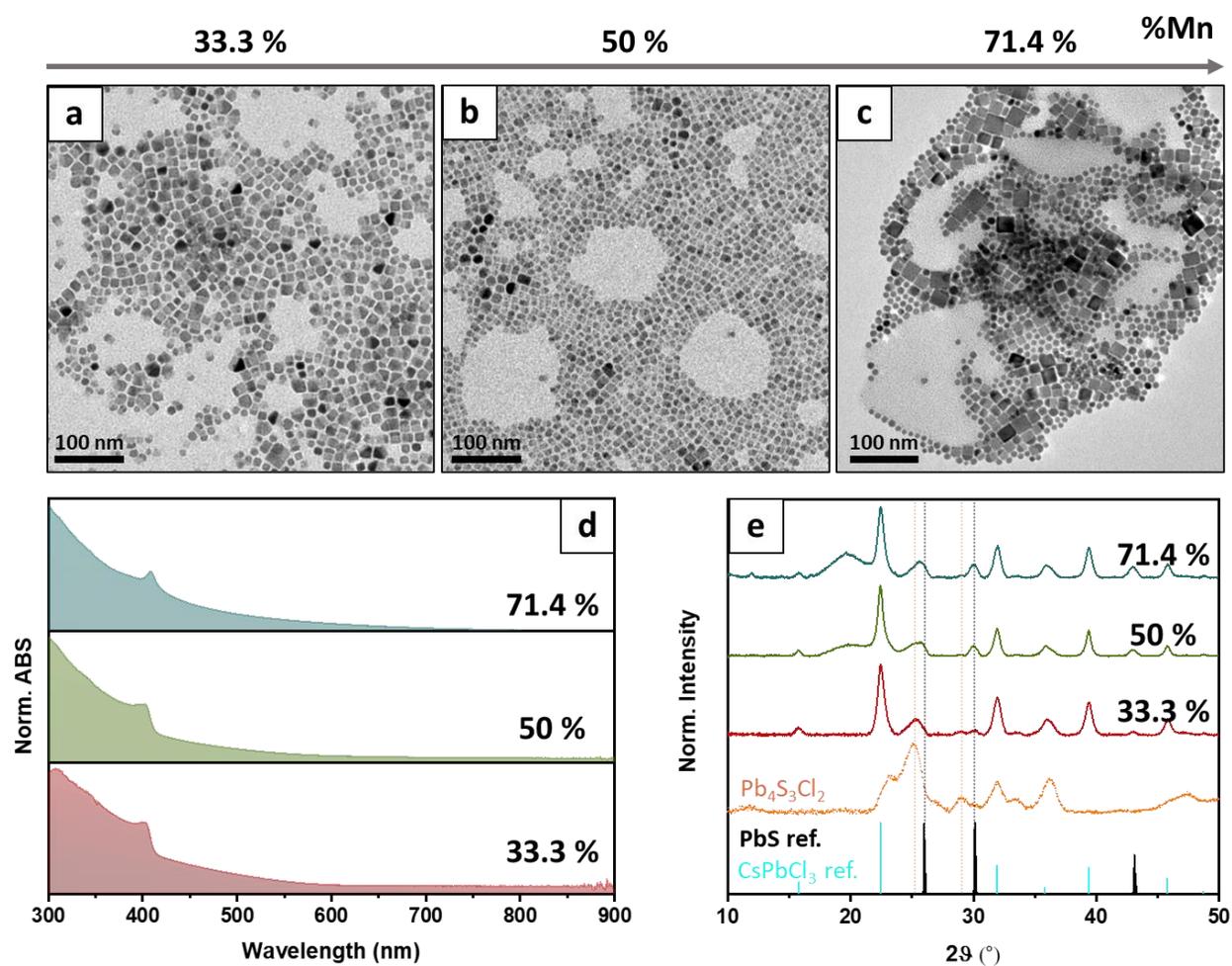

**Figure S4**. Influence of $Mn^{2+}$ feed ratio in the $CsPbCl_3$-PbS heterostructures synthesis (using Sulfur-ODE as a sulfur source) by varying the concentration of $Mn^{2+}$ added in the reaction system ($X_{Mn}$ = [Mn]/([Mn]+[Pb])*100). (a-c) TEM images and (d) optical absorption spectra of heterostructures obtained with different Mn-oleate precursor concentrations. (e) XRD patterns of $CsPbCl_3$-PbS heterostructures obtained with different $Mn^{2+}$ concentrations (blue, green, and red) and comparison with $Pb_4S_3Cl_2$ chalcohalides diffraction pattern (orange) and $CsPbCl_3$, and PbS reference patterns (cyan, and black respectively).

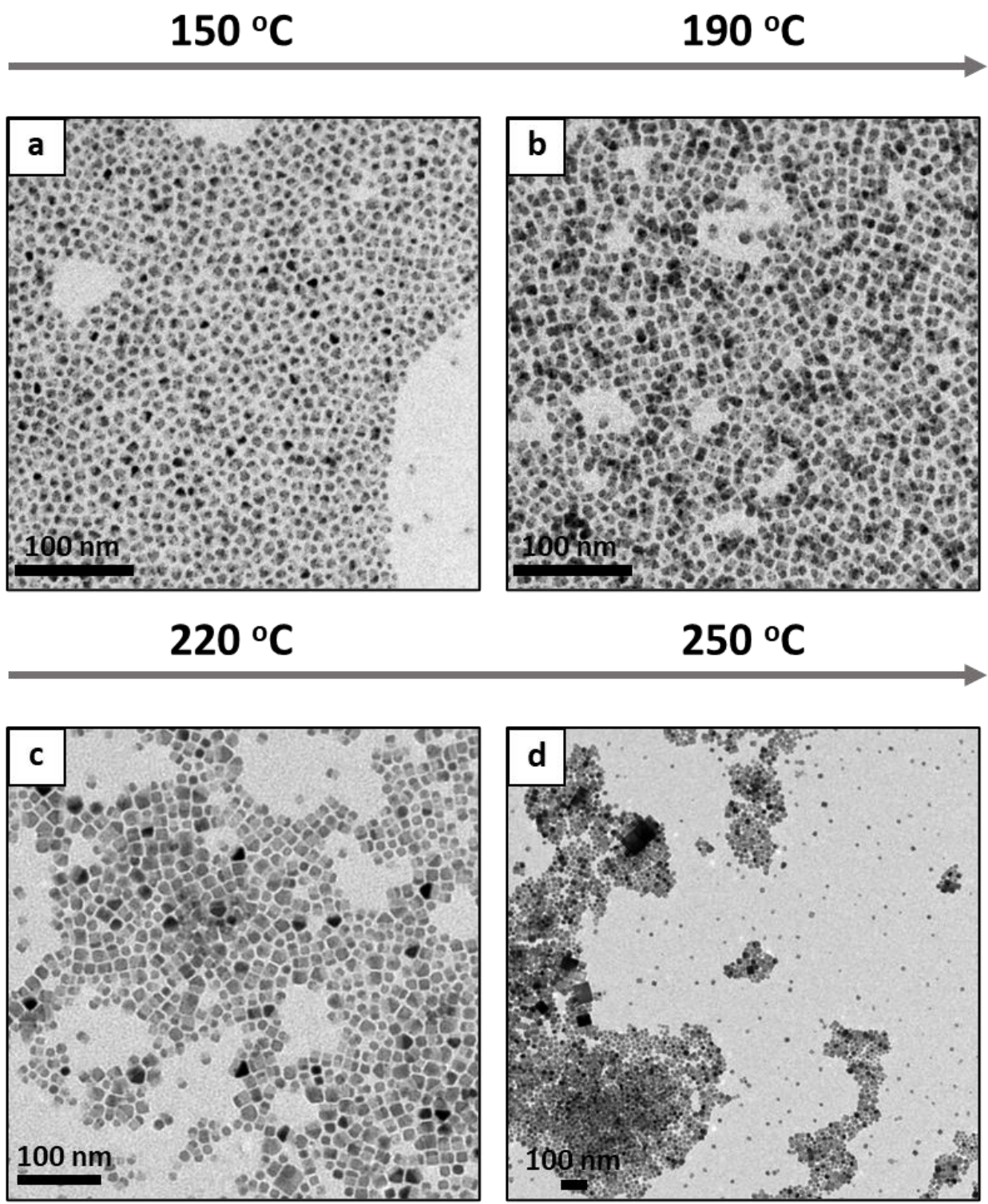

**Figure S5.** Influence of the reaction temperature on the formation of CsPbCl$_3$-PbS heterostructures using S-ODE as a sulfur source. TEM images of the heterostructures product were obtained at different reaction temperatures (150 to 250 °C) in the S-ODE synthesis and a reaction time of 5 min.

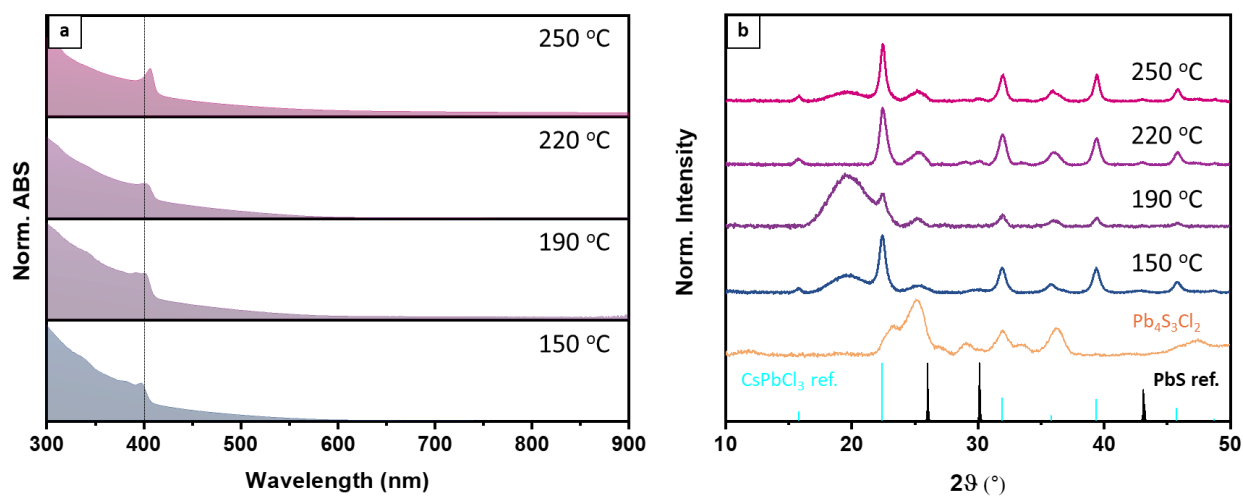

**Figure S6.** Influence of reaction temperature on the formation of CsPbCl$_3$-PbS heterostructures using S-ODE as a sulfur source. Reaction time is constant at 5 min. (a) UV-Vis-NIR absorption spectra of CsPbCl$_3$-based heterostructures obtained at different reaction temperatures and using S-ODE as a sulfur source. (b) XRD patterns of CsPbCl$_3$-based heterostructures obtained at different reaction temperatures and using S-ODE as a sulfur source and comparison with Pb$_4$S$_3$Cl$_2$ chalcohalides diffraction pattern (orange) and CsPbCl$_3$, and PbS reference patterns (cyan and black, respectively). The formation of the CsPbCl$_3$-PbS heterostructures population with the temperature is evidenced by the appearance of diffraction peaks at ~ 30º and 43º. The strong reflections attributed to the chalcohalide phase are representative of the dominant CsPbCl3-Pb$_4$S$_3$Cl$_2$ populations observed in corresponding TEM images in Figure S2.

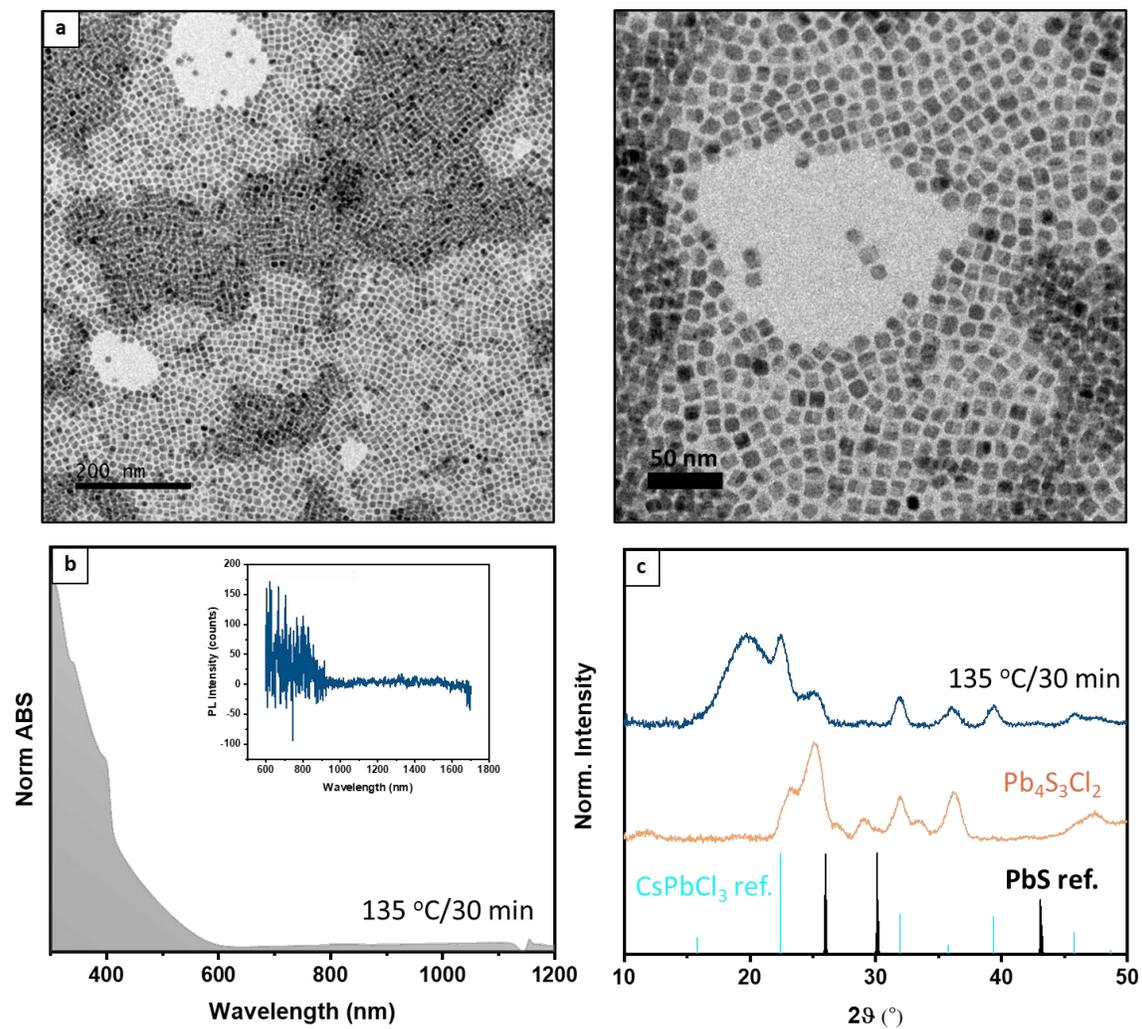

**Figure S7.** Influence of low reaction temperature (135 °C) on the formation of CsPbCl$_3$-PbS heterostructures using sulfur-ODE as a sulfur source. The reaction time is 30 min. (a) TEM images, (b) UV-Vis-NIR absorption and PL (inset) spectra, and (c) XRD patterns of the heterostructures product obtained at low temperatures (135 °C). XRD reference patterns are the following: orange for Pb$_4$S$_3$Cl$_2$, cyan for CsPbCl$_3$, and black for PbS.

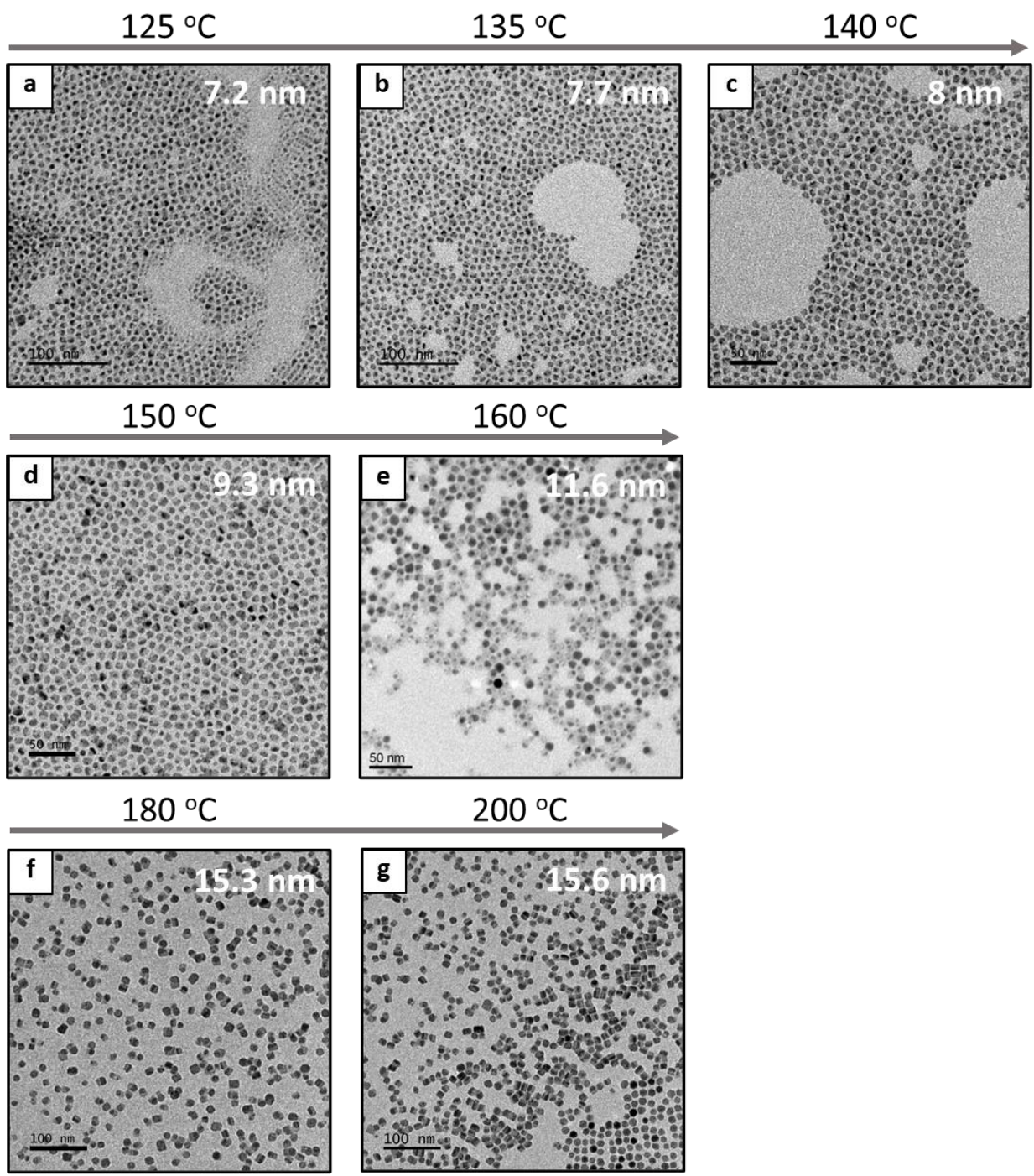

**Figure S8.** Influence of the reaction temperature on the formation of CsPbCl$_3$-PbS heterostructures using TMS-ODE as a sulfur source. The reaction time is 5 min. TEM images of CsPbCl$_3$-PbS heterostructures were obtained at different reaction temperatures (125 – 200 ºC). The images evidenced an increase in the heterostructure size with the reaction temperature. At temperatures higher than 160 °C, the heterostructures were more heterogeneous in size and shape.

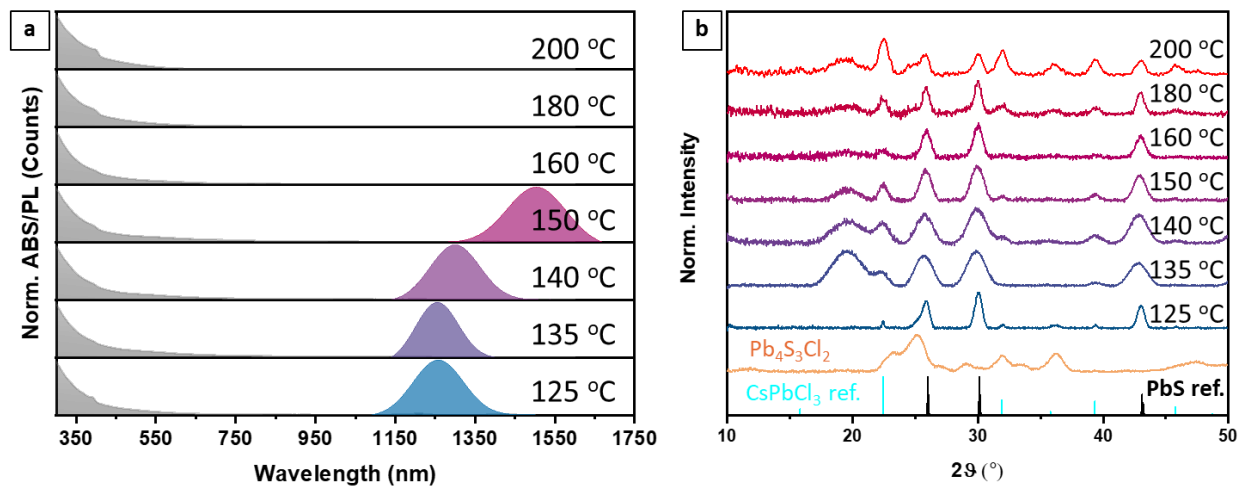

**Figure S9.** Influence of reaction temperature on the formation of the CsPbCl$_3$-based heterostructures using TMS-ODE as a sulfur source. Reaction time is constant at 5 min. (a) UV-Vis-NIR absorption and PL spectra, and (b) XRD patterns of CsPbCl$_3$-based heterostructures obtained at different reaction temperatures (125 – 200 °C). The XRD reference patterns (bottom) are the following: orange for Pb$_4$S$_3$Cl$_2$ chalcohalides, cyan for CsPbCl$_3$, and black for PbS. For reaction temperatures over 180 °C, the presence of Pb$_4$S$_3$Cl$_2$ chalcohalides as subproduct is evidenced by the appearance of a diffraction peak at ~ 25°.

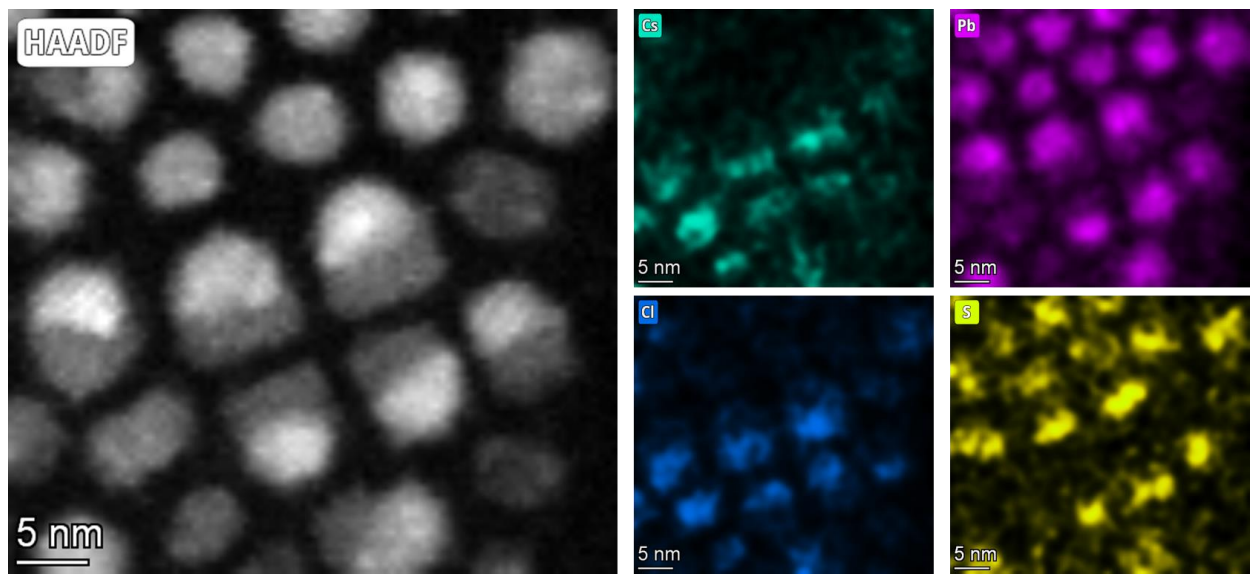

**Figure S10.** HAADF STEM image of CsPbCl$_3$-PbS heterostructures (case sample "125 °C/TMS", as described in Figure 1c) with the corresponding EDX elemental maps for cesium, lead, chlorine, and sulfur.

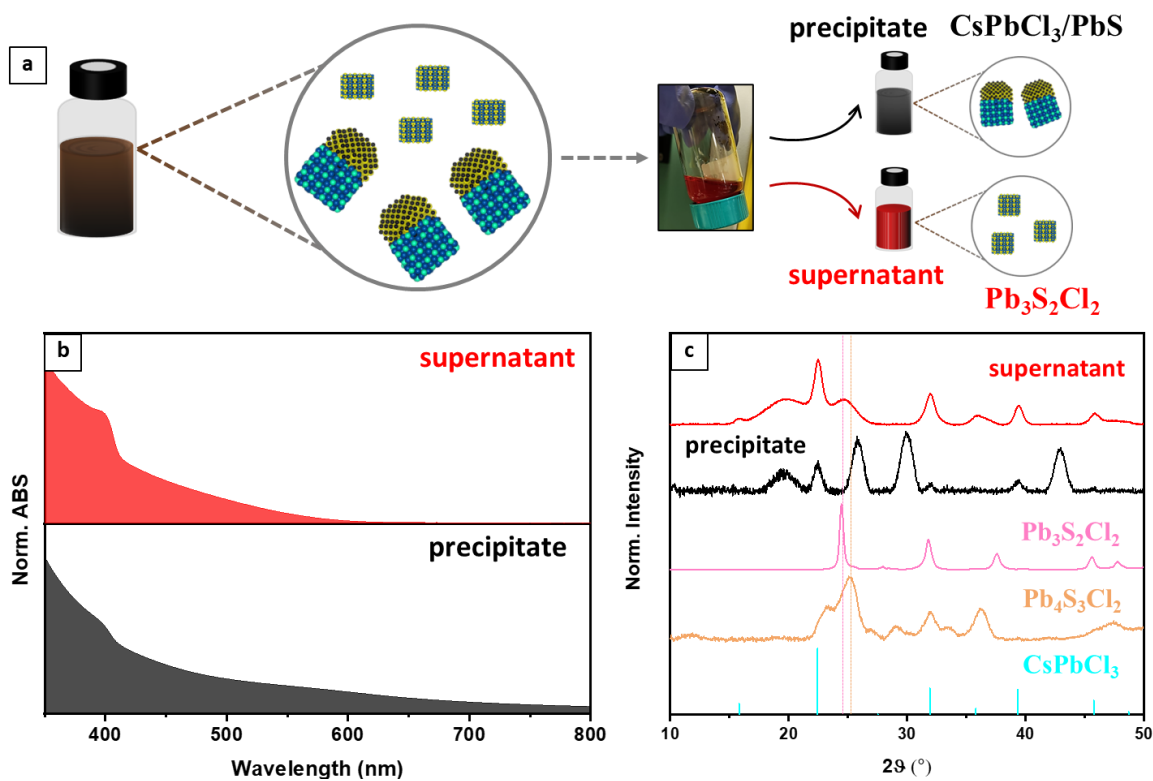

**Figure S11.** (a) Sketch of the purification process of a crude solution obtained at a reaction temperature of 150 °C using TMS as the sulfur source (case sample "150 °C/TMS", as described in Figure 1b). The 2 products obtained in the reaction ($CsPbCl_3$-PbS heterostructures and $Pb_3S_2Cl_2$ chalcohalides) are easily separated via centrifugation, remaining the $Pb_3S_2Cl_2$ chalcohalides in the supernatant (red) and precipitating the PbS based heterostructures (black). (b) UV-Vis-NIR absorption spectra of the supernatant and precipitate resulted from the purification process described in (a). The optical absorption in the 400-600 nm range indicates the presence of $Pb_3S_2Cl_2$ chalcohalides in the supernatant (top spectrum) accompanied by the absorption of $CsPbCl_3$ perovskite NCs. The absorption in the NIR range in the bottom spectrum with the excitonic peak at ~ 400 nm evidences the presence of $CsPbCl_3$-PbS heterostructures. (c) XRD patterns of the supernatant and precipitate resulted from the purification process described in (a) and comparison with $Pb_3S_2Cl_2$ chalcohalides diffraction pattern (pink), $Pb_4S_3Cl_2$ chalcohalides diffraction pattern (orange) and $CsPbCl_3$ reference pattern (cyan). The diffraction peak at ~ 24° indicates the presence of the $Pb_3S_2Cl_2$ phase in the supernatant (red). The absence of ~ 24° diffraction peak in the precipitate (black) confirms the $CsPbCl_3$-PbS phase purity.

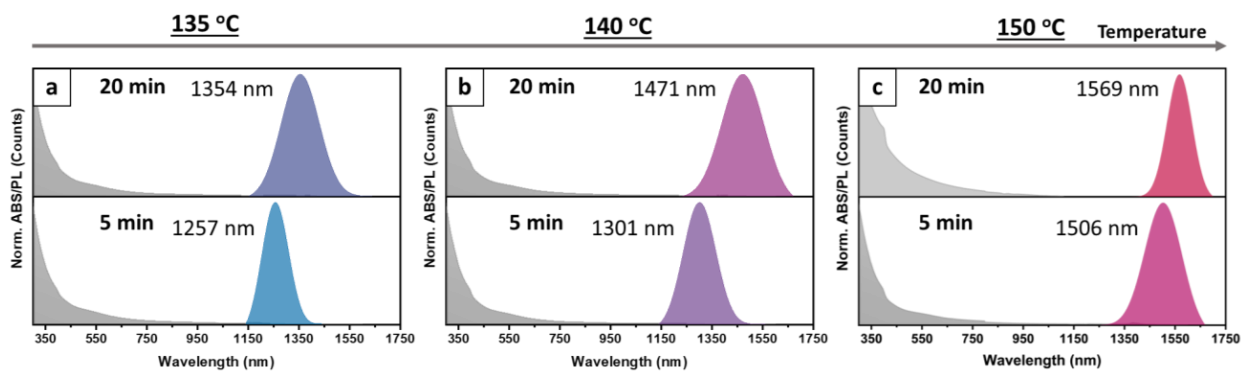

**Figure S12.** Influence of reaction temperature and reaction time on the growth of the $CsPbCl_3$-PbS heterostructures using TMS as a sulfur source. (a-c) UV-Vis-NIR absorption and PL spectra of $CsPbCl_3$-PbS heterostructures obtained at different reaction temperatures: 135 °C (a), 140 °C (b), and 150 °C (c); and different reaction times: 5 min (bottom spectra), and 20 min (top spectra).

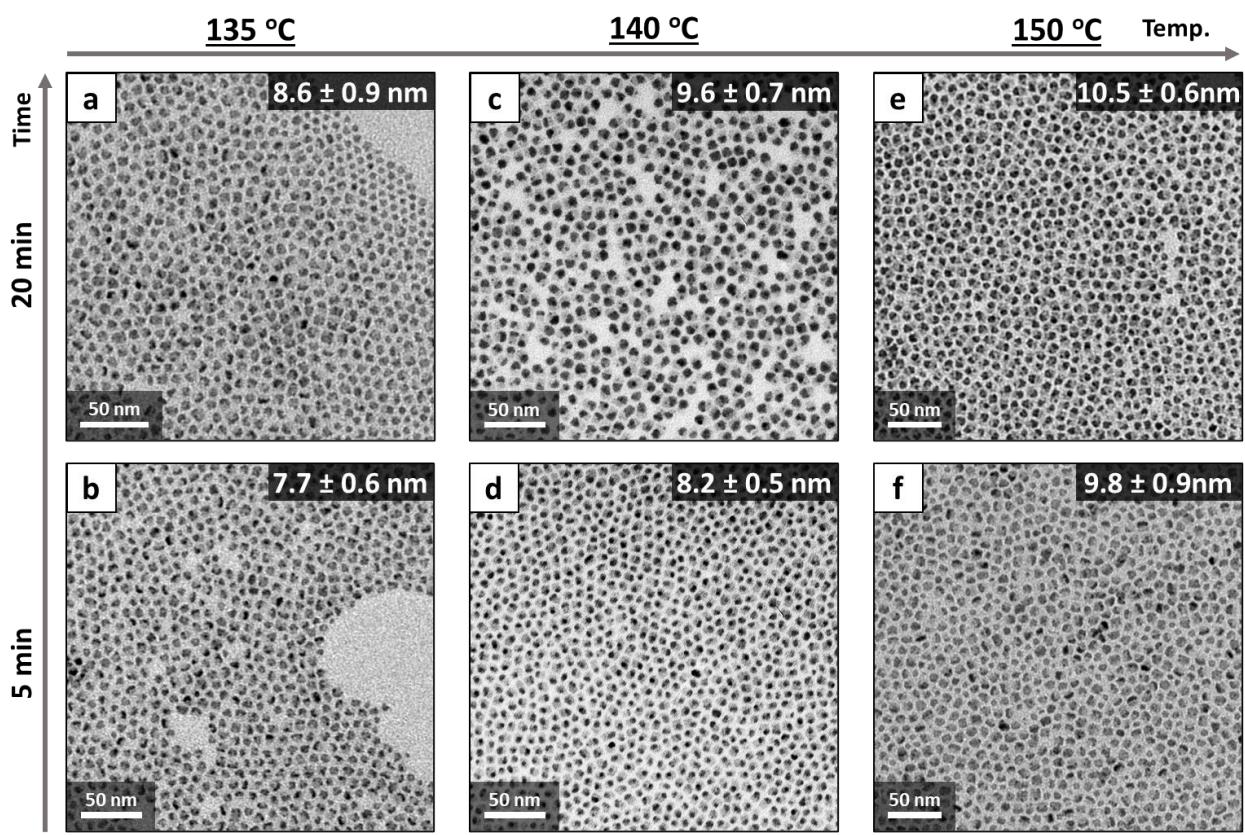

**Figure S13**. Influence of reaction temperature and reaction time on the growth of the CsPbCl$_3$-PbS heterostructures using TMS as a sulfur source. (a-f) TEM images of CsPbCl$_3$-PbS heterostructures obtained at different reaction temperatures: 135 °C (a,b), 140 °C (c,d), and 150 °C (e,f); and different reaction times: 5 min (bottom images), and 20 min (top images).

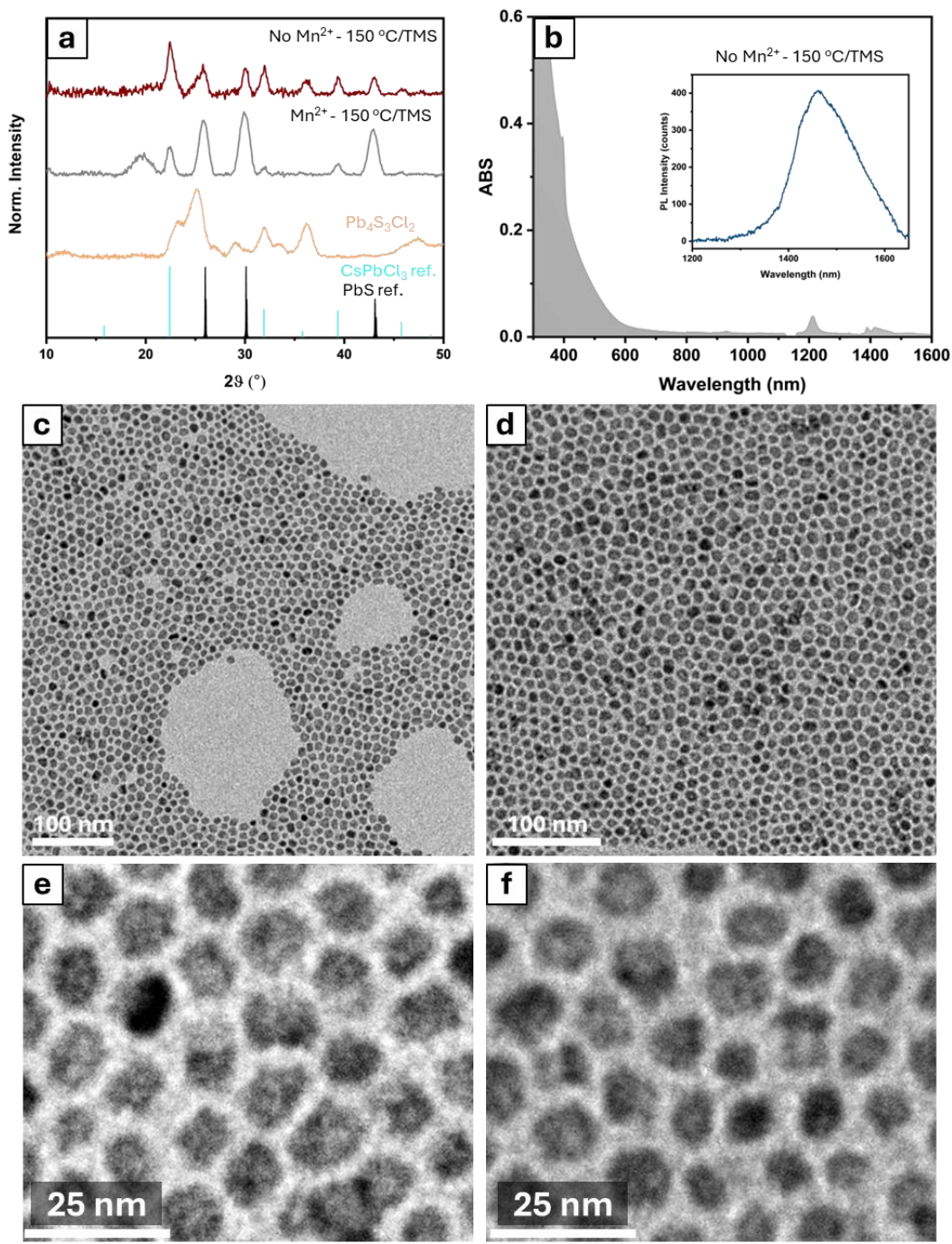

**Figure S14.** Analysis of the synthesis of case sample "150 °C/TMS" in the absence of $Mn^{2+}$. (a) XRD diffraction patterns of the products synthesized via the "150 °C/TMS" method with (grey) and without (red) Mn-OL, and comparison with the $Pb_4S_3Cl_2$ chalcohalides diffraction pattern (orange) and $CsPbCl_3$, and PbS reference patterns (cyan, and black, respectively). The absence of $Mn^{2+}$ in the reactions yields a product containing both $Pb_4S_3Cl_2$ and PbS phases (along with the perovskite) as indicated by the peaks at 25.2° and 26.2°, respectively. (b) Optical absorption and PL spectra of the sample synthesized with "150 °C/TMS" in the absence of Mn-OL. (c,d) TEM images of the same sample demonstrate a mixture of productcs in addition to the $CsPbCl_3$-PbS heterostructures. Therefore, it is likely that the trimethylsilyl group of TMS can act as an effective $Cl^-$ scavenger. (e,f) Enlarged areas of the TEM image presented in (c).

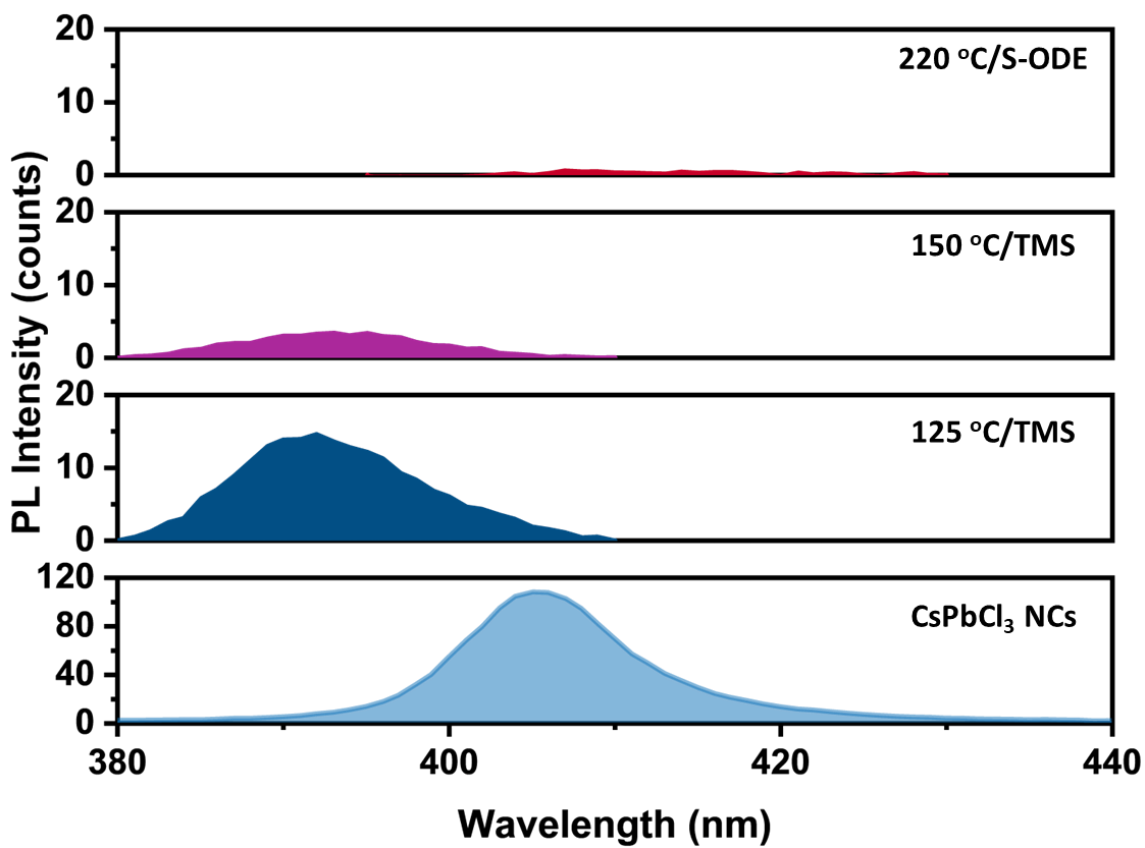

**Figure S15**. Photoluminescence spectra in the visible range capturing the emission originated from the perovskite domain (λexc=350 nm). The four spectra correspond to pure $CsPbCl_3$ NCs and the three cases/samples of heterostructures represented in Figure 1. Notably, the $CsPbCl_3$ NCs were synthesized based on the clusters-based approach used for the synthesis of the heterostructure. This suggests similar surface capping (passivation) between the pure perovskite NCs and heterostructures. In every case, the perovskite emissions are severely quenched when they are bound to the PbS or $Pb_4S_3Cl_2$ domains.

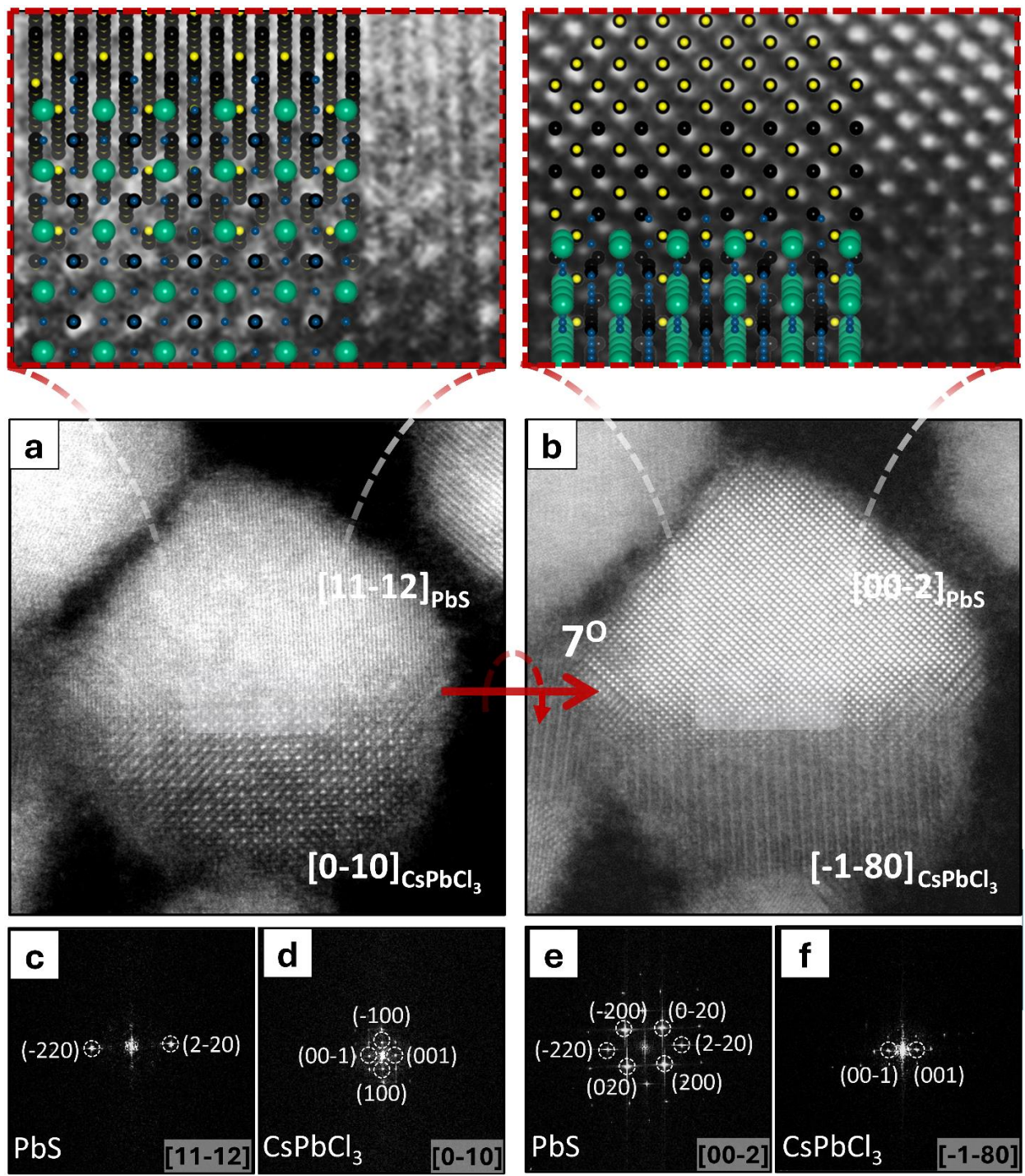

**Figure S16**. **Structural and compositional analysis of CsPbCl$_3$/PbS epitaxial heterostructures.** (a-b) HAADF STEM images of the same heterostructure tilted by 7° around the x-axis to align both domains on low-index zone axes. (c,d), and (e,f) represent FFT patterns of the image a, and b, respectively. For each image, we present two FFTs corresponding to the two distinct domains of the heterostructure. (a,b upper panels) represent magnified areas of the STEM images. The constructed 3D model fits the atomic columns in this projection as well, validating the model. FFT patterns (c,d, or e,f) suggest that the zone axes of the heterostructure projection presented in Figure 2b are parallel planes in this orthogonal heterostructure projection presented here (a or b).

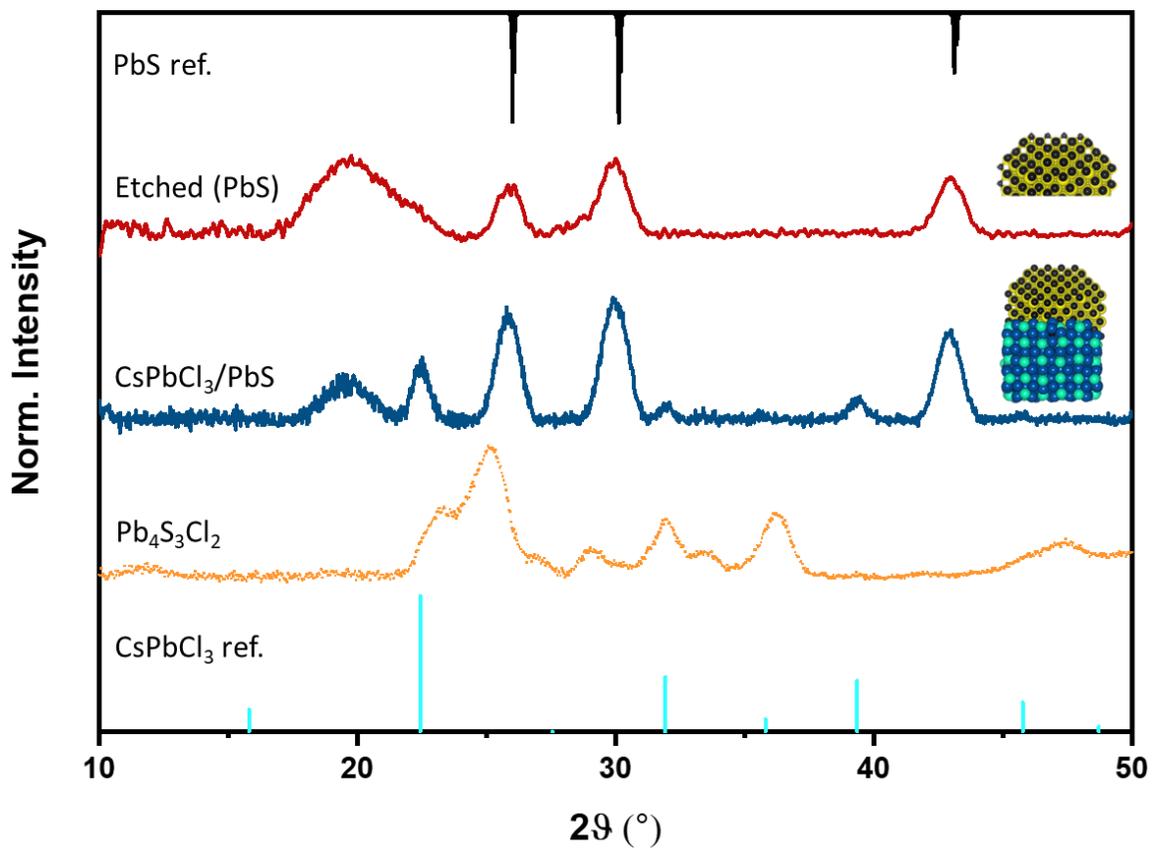

**Figure S17.** XRD patterns of CsPbCl$_3$/PbS heterostructures (sample "150 °C/TMS" corresponding to the blue pattern), and the etched CsPbCl$_3$/PbS heterostructures (thus, PbS NCs represented in the red pattern). Additional XRD patterns are displayed for comparison: Experimental pattern of Pb$_4$S$_3$Cl$_2$ (orange), and reference patterns of CsPbCl$_3$ (cyan lines), and PbS (black lines). The absence of reflections at 22.6°, 32°, and 39.5° in the etched sample, confirms the successful etching of the perovskite domain, and thus obtaining pure PbS NCs. The broad peak at approximately 19° is attributed to the fumed silica used for performing the XRD measurements.

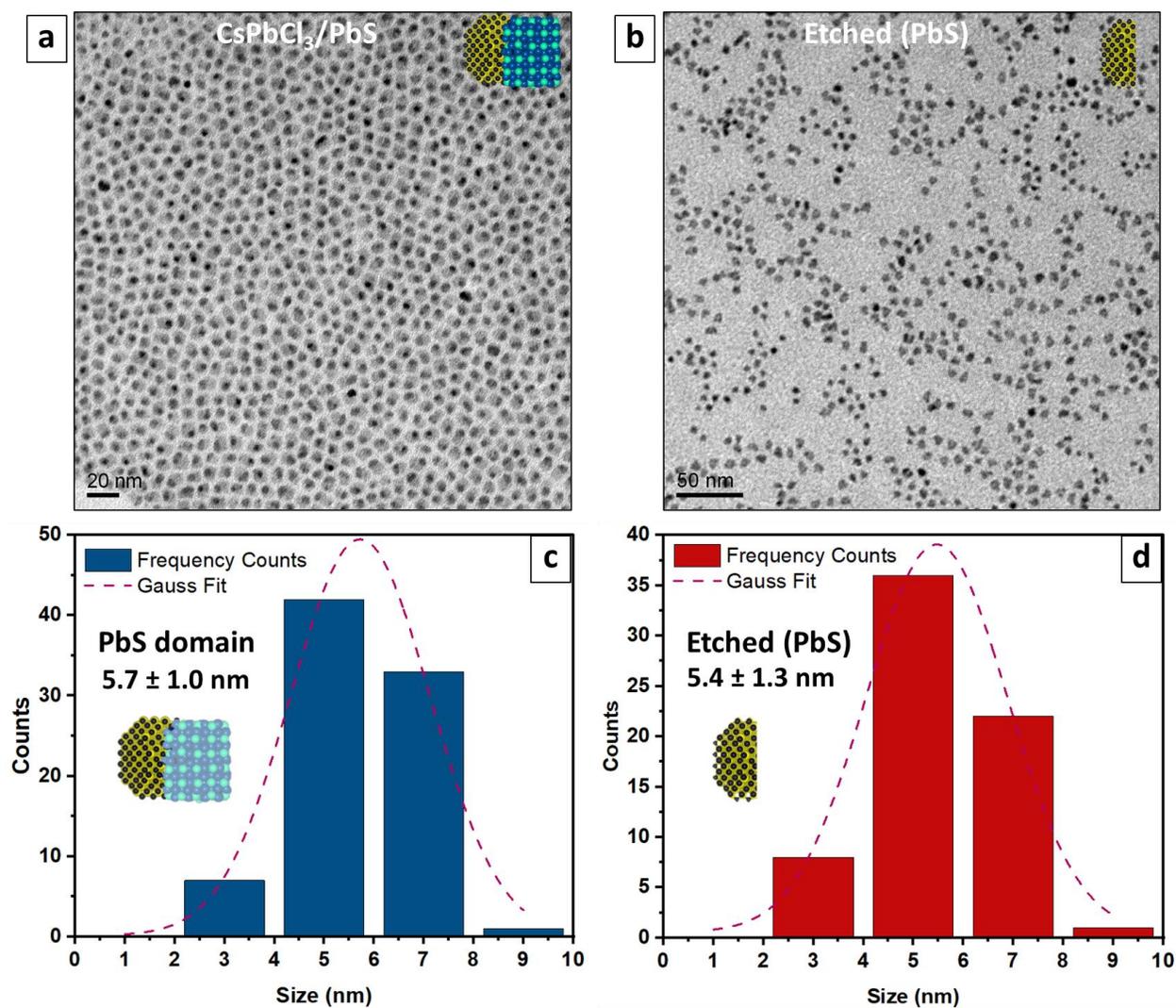

**Figure S18.** Morphological comparison of CsPbCl₃/PbS heterostructures (sample "150 °C/TMS"), and the etched CsPbCl₃/PbS heterostructures (thus, PbS NCs). (a,b) Low magnification TEM images of CsPbCl₃/PbS heterostructures (panel a), and the etched CsPbCl₃/PbS heterostructures (i.e. PbS NCs, panel b). TEM images ensure the successful etching of the perovskite domain and thus, obtain pure PbS NCs, as already indicated by the XRD patterns (Figure S13) and optical absorption spectra (Figure 3d). (c,d) Size distribution histograms of the PbS domain in the case of the heterostructures (panel c), and after etching (panel d) comparing the size difference of the PbS before and after the etching. A slightly smaller size distribution of the PbS domain is revealed after etching. However, this difference is within the margin of error. The NC size distribution histograms were estimated using Ilastik software.

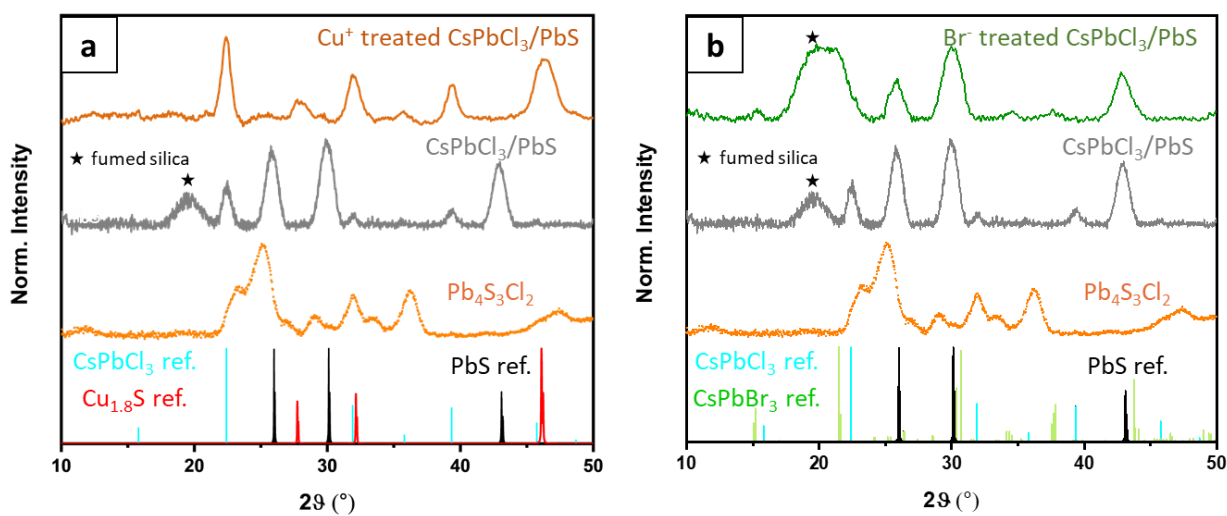

**Figure S19.** XRD patterns of cation ($Pb^{2+} \rightarrow Cu^+$) exchanged $CsPbCl_3/Cu_{2-x}S$ (a), and anion ($Cl^- \rightarrow Br^-$) exchanged $CsPbBr_3/PbS$ heterostructures (b). The parent sample of $CsPbCl_3/PbS$ heterostructures, used for these exchanges was defined in the text as "150 °C/TMS", with the corresponding XRD pattern presented in both panels for comparisons (grey color). (a) The $Cu^+$ treated $CsPbCl_3/PbS$ heterostructures reveal full cation exchange to $CsPbCl_3/Cu_{2-x}S$ heterostructures. This is confirmed by the absence of PbS reflection peaks in the exchanged sample (at 30.2°, and 43.1°). The sub-stoichiometric $Cu_{2-x}S$ phase is ascribed to the composition of the digenite ($Fm\bar{3}m$) $Cu_{1.8}S$ phase (reference pattern represented with red lines) based on the position of the reflection peaks. The broad peak (asterisk) at approximately 19° is attributed to the fumed silica used for performing XRD measurements. Additional XRD patterns are demonstrated for comparison: Experimental pattern of $Pb_4S_3Cl_2$ (orange), and reference patterns of $CsPbCl_3$ (cyan lines), and PbS (black lines). (b) The $Br^-$ treated $CsPbCl_3/PbS$ heterostructures reveal full anion exchange to $CsPbBr_3/PbS$ heterostructures. Notably, the dominant peak of the $CsPbBr_3$ orthorhombic phase overlaps with the fumed silica (~21.5°). The reference pattern of $CsPbBr_3$ (green lines) is also shown.

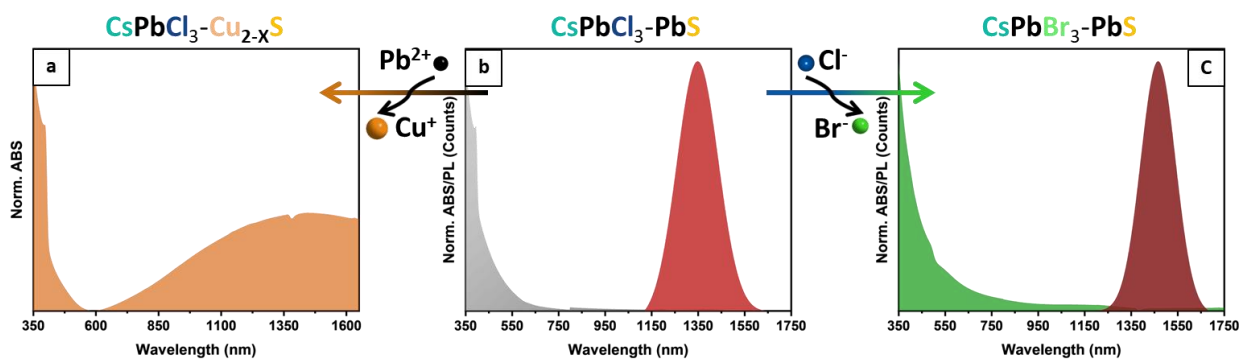

**Figure S20.** Optical absorption spectra of pure $CsPbCl_3/PbS$ (center), cation ($Pb^{2+} \rightarrow Cu^+$) exchanged $CsPbCl_3/Cu_{2-x}S$ (left), and anion ($Cl^- \rightarrow Br^-$) exchanged $CsPbBr_3/PbS$ heterostructures (right). The parent sample of $CsPbCl_3/PbS$ heterostructures, used for these exchanges was defined in the text as "150 °C/TMS". These are the same samples that are described in Figure S15.

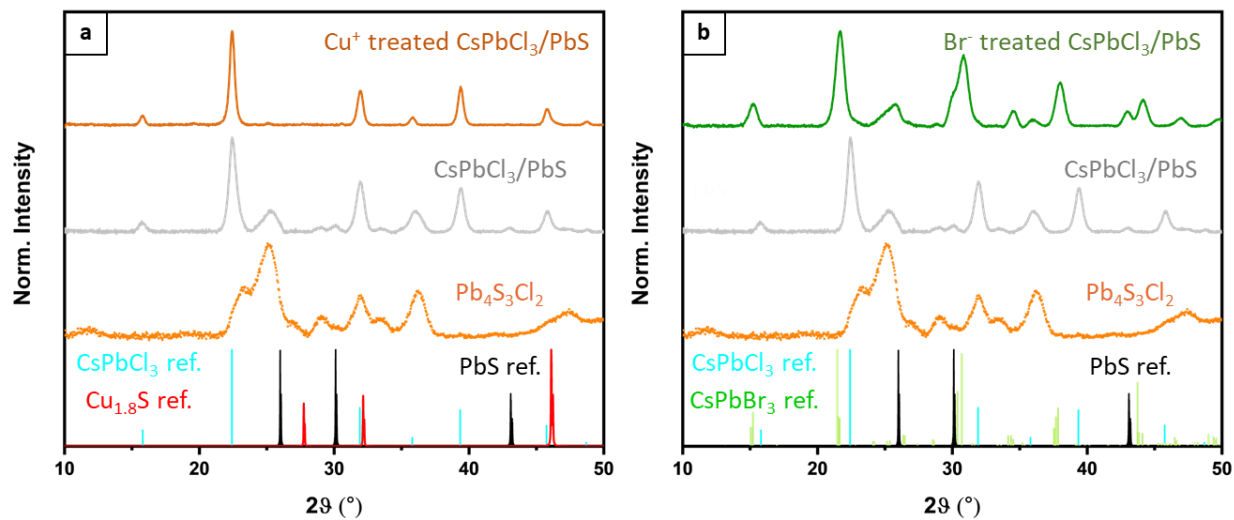

**Figure S21.** XRD patterns of cation ($Pb^{2+}\rightarrow Cu^+$) exchanged $CsPbCl_3/Cu_{2-x}S$ (a), and anion ($Cl^-\rightarrow Br^-$) exchanged $CsPbBr_3/PbS$ heterostructures (b). The parent $CsPbCl_3/PbS$ sample, used for these exchanges, was defined in the text as "220 °C/S-ODE". The XRD pattern of the parent sample is presented in both panels for comparison (grey color).

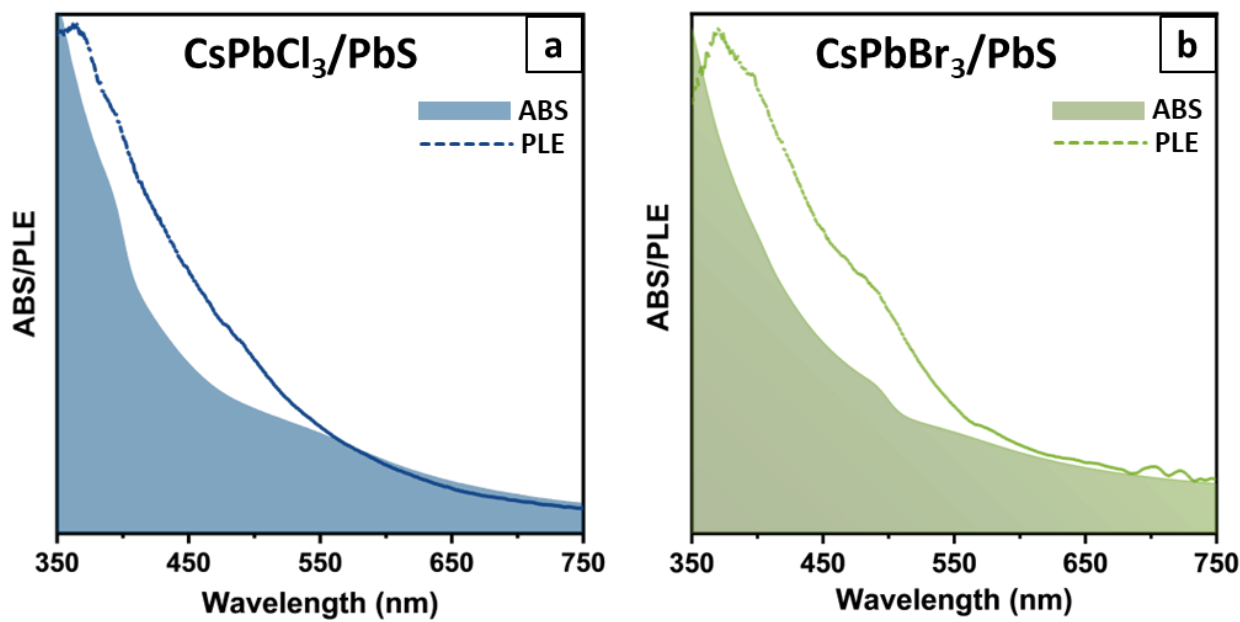

**Figure S22.** Optical absorption (colored curves) and PL excitation (dashed lines) spectra for both (a) $CsPbCl_3/PbS$, and anion exchanged (b) $CsPbBr_3$-PbS heterostructures (sample "150 °C/TMS"). PLE spectrum of the $CsPbBr_3/PbS$ (emission at 1470 nm) closely resembles the corresponding absorption spectrum. The peak at ~ 500 nm, which relates to the perovskite feature is more pronounced than in the case (a) of $CsPbCl_3/PbS$ heterostructures. This difference could be due to the reduced contribution from absorption by PbS at the $CsPbBr_3$ exciton peak. This observation is in agreement with the Type-I alignment suggesting that photoexcited carriers in the $CsPbBr_3$ domain are transferred to the PbS domain.

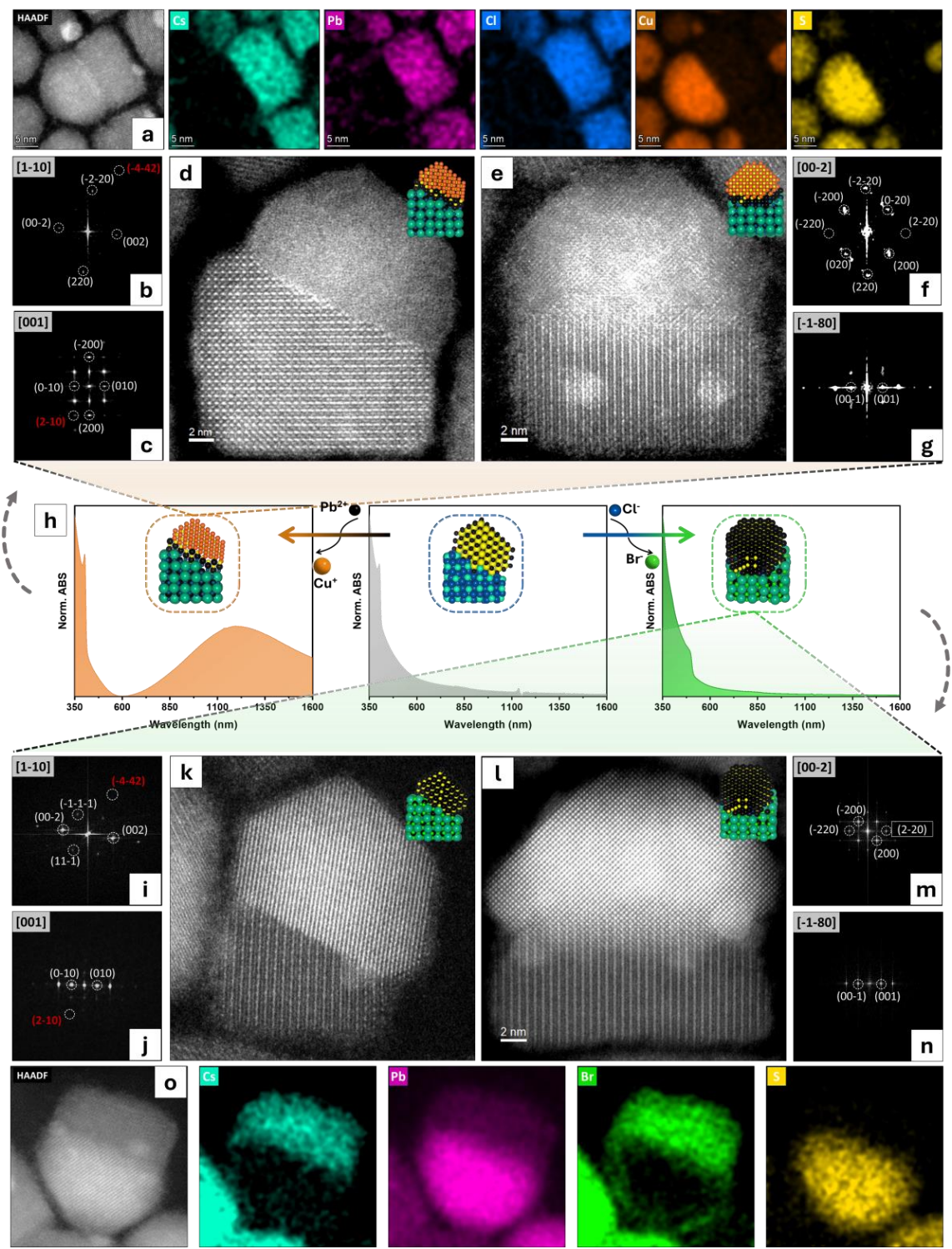

**Figure S23. Phase selective cation ($Pb^{2+}\rightarrow Cu^+$) and anion ($Cl^-\rightarrow Br^-$) exchanges on $CsPbCl_3$/PbS heterostructures.** (a) HAADF STEM image of a single $CsPbCl_3$/$Cu_{2-x}S$ heterostructure with the corresponding STEM-EDX elemental maps. (d,e) HAADF HR STEM images of two different cation-exchanged $CsPbCl_3$/$Cu_{2-x}S$ heterostructures in common orthogonal projections. (b,c) FFT patterns of $Cu_{2-x}S$ (b) and $CsPbCl_3$ (c) domains of the HAADF STEM image are depicted in panel d. Red dashed circles in the FFTs in panels (b) and (c) represent planes that are parallel to the interface of the heterostructure, as described in Figure 2 for the parent $CsPbCl_3$/PbS heterostructures. (f,g) FFT patterns of $Cu_{2-x}S$ (f) and $CsPbCl_3$ (g) domains of the HAADF STEM image in panel (e). FFT patterns demonstrate that the zone axes of each domain of the first projection (d) are parallel planes in the second projection (e), ensuring orthogonality (see blue dashed spots in panels f and g). (h) Optical absorption spectra of pure $CsPbCl_3$/PbS (center), cation ($Pb^{2+}\rightarrow Cu^+$) exchanged $CsPbCl_3$/$Cu_{2-x}S$ (left), and anion ($Cl^-$

→Br⁻) exchanged CsPbBr₃/PbS heterostructures (right). (k-l) HAADF HR STEM images of two different anion-exchanged CsPbBr₃/PbS heterostructures in orthogonal projections. (i,j) FFT patterns of PbS (i) and CsPbBr₃ (j) domains of the HAADF STEM image depicted in panel k. Red dashed circles in the FFTs represent planes that are parallel constituting the interface of heterostructure. (m,n) FFT patterns of PbS (m) and CsPbBr₃ (n) domains of the HAADF STEM image in panel l. (o) HAADF STEM image of a single CsPbBr₃/PbS heterostructure with the corresponding STEM-EDX elemental maps.

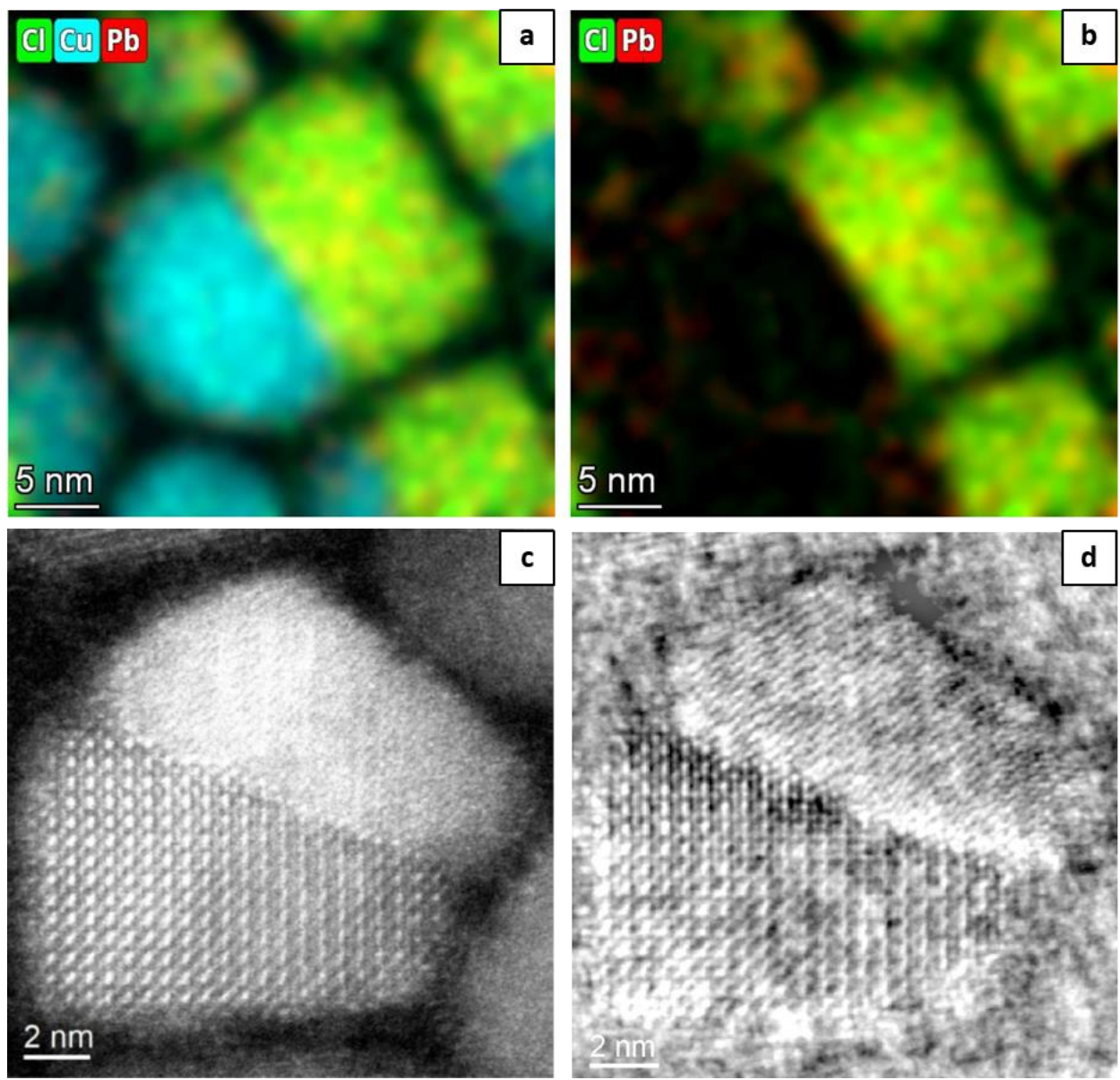

**Figure S24**. (a,b) STEM-EDX elemental maps of a single CsPbCl$_3$/Cu$_{2-x}$S heterostructure, indicating a non-exchanged monolayer of PbS at the interface (where Cl⁻ ions are not detected, indicating that this Pb layer does not belong to the perovskite domain). (c) HAADF HR STEM-image of cation exchanged CsPbCl$_3$/Cu$_{2-x}$S. (d) iDPC STEM image of the same area.

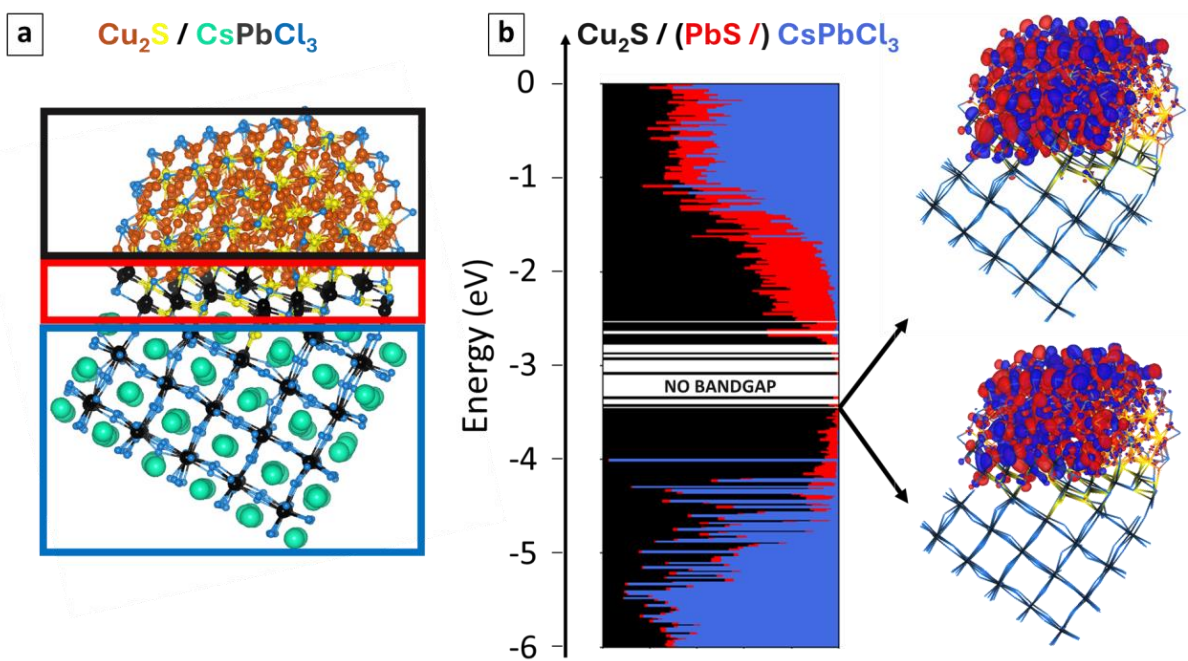

**Figure S25.** (a) Ball and sticks representation of the CsPbCl$_3$/Cu$_2$S NC heterostructure model optimized at the DFT/PBE level of theory: the presence of an intermediate PbS layer results in a smooth transition between the two domains as it allows to complete the coordination of the ions of both CsPbCl$_3$ and Cu$_2$S domains at the interface. (b) Electronic structure of the model represented in (a) computed at the DFT/PBE level of theory. The color code indicates the contribution of each domain to each molecular orbital. On the right, we plotted the frontier molecular orbitals. The absence of bandgap is most likely due to the presence of surface defects and to an underestimation of the band gap typical of some of the DFT exchange-correlation functionals, like the PBE employed here.